\documentclass[aps, prl, twocolumn, showpacs]{revtex4-1}
\usepackage{amsmath}
\usepackage{amssymb}
\usepackage{bm}
\usepackage{graphicx}
\usepackage[colorlinks=true, allcolors=blue]{hyperref}
\usepackage{float}
\usepackage[normalem]{ulem}
\usepackage{xcolor}    % 必须加

\usepackage{dcolumn}
\usepackage{tikz}
\usepackage[compat=1.1.0]{tikz-feynman}
\immediate\write18{texcount -inc -sum main.tex > wordcount.txt}

\begin{document}
\setlength{\parindent}{2em}

\title{False Vacuum Decay across the Quantum-to-Thermal Crossover: A Comparison of Real-Time Observables}

\author{Haiyang Wang}
\affiliation{Department of Physics and Chongqing Key Laboratory for Strongly Coupled Physics, Chongqing University, Chongqing 401331, P. R. China}

\author{Renhui Qin}
\affiliation{Department of Physics and Chongqing Key Laboratory for Strongly Coupled Physics, Chongqing University, Chongqing 401331, P. R. China}

\author{Ligong Bian}
\email{lgbycl@cqu.edu.cn}
\affiliation{Department of Physics and Chongqing Key Laboratory for Strongly Coupled Physics, Chongqing University, Chongqing 401331, P. R. China}

\begin{abstract}   
We develop a real-time Wigner-functional lattice framework with positive Hartree-Gaussian initial sampling and introduce a connected-cluster survival criterion for extracting false-vacuum decay rates across the crossover from quantum fluctuations to thermal nucleation. At high temperatures, the connected-cluster rate agrees well with the Hartree-resummed thermal nucleation benchmark, while the commonly used global-survival criterion can give substantially smaller rates because of multi-seed dynamics and global averaging. At low temperatures, the connected-cluster and global-survival rates approach each other in the dilute-event regime, whereas the false-vacuum fraction observable can be contaminated by transient spatial conversion and kink-antikink reflection. Our results clarify how different real-time observables encode distinct aspects of metastable decay.

\end{abstract}

\maketitle

{\it Introduction.---}
False vacuum decay (FVD) can occur through quantum tunneling at zero temperature~\cite{Coleman:1977py,Callan:1977pt} and/or through thermal fluctuations in the early universe~\cite{Linde:1981zj}.
Quantum tunneling is crucial for understanding the decay rate of the electroweak vacuum or the age of our universe~\cite{Chigusa:2017dux,Andreassen:2017rzq}. The foundational quantum field theory description of quantum tunneling was established in Refs.~\cite{Coleman:1977py,Callan:1977pt}, and was later extended to finite-temperature scenarios to consider the thermal dynamical fluctuations~\cite{Linde:1981zj,Affleck:1980ac} based on the functional integral treatment of classical metastability~\cite{Langer:1967ax,Langer:1969bc},which is closely related to Kramers' escape rate theory in the field theory~\cite{Berera:2019uyp}. Thermal fluctuations can generate localized field configurations, including oscillons, which may seed the false vacuum decay through vacuum bubble nucleation~\cite{Copeland:1995fq,Gleiser:1996jb,Gleiser:1993pt,Gleiser:2004iy,Gleiser:2007ts,Pirvu:2023plk}. Vacuum bubble dynamics in the first-order cosmological thermal phase transition can generate primordial magnetic fields~\cite{Di:2025ncl,Di:2024gsl,Liu:2024mdo,Yang:2021uid,Di:2020kbw,Ahonen:1997wh,Stevens:2012zz,Zhang:2019vsb}, detectable gravitational waves at space-based interferometers (such as LISA~\cite{LISA:2017pwj}, TianQin~\cite{Luo:2025ewp} and Taiji~\cite{Bian:2025ifp,Ruan:2018tsw}), and provide a possible mechanism for explaining the baryon asymmetry of the universe through electroweak baryogenesis~\cite{Cohen:1993nk,Morrissey:2012db,Trodden:1998ym}.

It is known that the FVD rates (bubble nucleation rates) in both the quantum tunneling and thermal fluctuation scenarios can be calculated based on the saddle-point or the instanton approximation, 
leaving the prefactor theoretically difficult to compute and the real-time dynamics of the decay process largely inaccessible. Drawing an analogy from quantum mechanics, Refs.~\cite{PhysRevLett.117.231601,PhysRevD.95.085011} developed a direct method for calculating the FVD at zero temperature.
More recent progress has shown that FVD rates in the quantum tunneling regime can be computed using real-time semiclassical simulations, in which the truncated Wigner method incorporates quantum corrections through vacuum fluctuations in the initial state, while the subsequent classical evolution allows bubble nucleation to occur~\cite{Pirvu:2021roq,PhysRevLett.123.031601}, which was found to be largely different from the prediction by the instanton method~\cite{Hertzberg:2020tqa}\footnote{To reduce the discrepancy between the two methods, Ref.~\cite{Braden:2022odm} suggests including lattice-renormalization effects in the instanton calculation.}.

In this Letter, we present a finite-temperature Wigner-functional lattice approach to FVD, which provides a unified real-time description  of FVD across both quantum and thermal regimes, overcoming the long-standing limitation of conventional saddle-point approximation methods. The Wigner-functional formalism provides a phase-space quasi-probability description of field configurations, with both quantum and thermal fluctuations systematically incorporated via the Hartree approximation. We then introduce two complementary observables to extract the FVD rate from real-time (1+1)-dimensional lattice simulations: connected-cluster method, which resolves microscopic individual bubble nucleation events, and the false vacuum fraction method, which captures the macroscopic nonlinear bubble growth dynamics. By benchmarking the FVD rates extracted from our two methods against the corresponding results obtained via the conventional global survival method widely used in the literature, we demonstrate that the connected-cluster method provides a robust sample-level diagnostic across the temperature range studied from the zero-temperature quantum tunneling limit, through the intermediate mixed quantum-thermal regime, to the high-temperature thermal nucleation regime. Finally, we employ a coarse-graining scheme to further suppress ultraviolet short-distance fluctuations and cleanly isolate the low-energy sector directly relevant to bubble nucleation. 

{\it Wigner functional approach to the false vacuum probability.---}
%In the classical case of a metastable system with two ground states, a potential barrier prevents a particle from entering one ground state from another when its energy is lower than the barrier. However, when quantum effects are taken into account, the particle can still enter one ground state from another. The probability distribution of the system can be calculated once the wave function expression has been obtained, the central quantity of interest is the probability that the configuration still belongs to the false vacuum basin at time $t$. However, in quantum field theory, it is difficult to obtain an analytic solution for the wave function in the interacting case. 
Since the Wigner function serves as a quasi-probability distribution, we extend this construction to scalar field theory~\cite{PhysRevD.50.7542}, and define the probability of the false vacuum as:
\begin{equation}
P_{FV}(t)= \frac{1}{Z}\int \mathcal{D}\phi \int \frac{\mathcal{D}\Pi}{2\pi}\,
\Theta_{FV}[\phi]\,
W[\phi,\Pi;t]
\label{PFV_with_wigner_new}
\end{equation}
where $\Theta_{FV}[\phi]$ is an indicator functional that returns unity if the field configuration $\phi(x)$ is identified as belonging to the false-vacuum basin, and vanishes otherwise.
Here $W[\phi,\Pi;t]$ is the Wigner functional and $Z$ is the normalization factor obtained from the same functional integral without the insertion of $\Theta_{FV}$.
%It is important to stress that Eq.~(\ref{PFV_with_wigner_new}) only defines the false vacuum probability. The subsequent kinetic law depends on how $\Theta_{FV}$ is implemented and on the physical observable extracted from the ensemble.
% In the following text, we will discuss this point in detail.
By analogy with the Wigner function in quantum mechanics, its definition is given by:
\begin{widetext} % 跨双栏公式
\begin{equation}
\label{W_def}
\begin{aligned}
W[\phi(x), \Pi(x) ; t]= \int \mathcal{D} \varphi(x) \exp \left[-\frac{i}{\hbar} \int d x \Pi(x) \varphi(x)\right] \times\left\langle\phi(x)+\frac{\varphi(x)}{2}\right| \hat{\rho}(t)\left|\phi(x)-\frac{\varphi(x)}{2} \right\rangle\;
\end{aligned}
\end{equation}
\end{widetext}
Here $\phi(x)$ is the time-independent real scalar field in the Schrödinger picture. $\Pi(x) \equiv \frac{\delta \mathcal{L}}{\delta \dot{\phi}(x)}$ is the  conjugate momentum density.
% $\hat{\rho}(t)=\left|\Psi(t)\right\rangle \left\langle \Psi(t)\right|$.
% We can use $\left\langle \phi(x) | \Psi(t) \right\rangle$ to obtain the wave function $\Psi(\phi, t)$, which is a functional of the field configuration and evolves with time. In quantum mechanics, once we know the wave function, we can obtain the probability that the system is in a certain interval in coordinate space. Therefore, we extend this idea to field theory, drawing an analogy in field configuration space, and obtain the probability that the field takes values within a given region.
Ref. \cite{PhysRevD.50.7542} shows that the Wigner function in QFT plays a role similar to that of a density distribution in the space spanned by $\phi$ and $\Pi$. In the Schr\"odinger picture, the density matrix operator $\hat\rho(t)$ satisfies the von Neumann equation:
\begin{equation}\label{rhoH}
    i \hbar  \frac{\partial}{\partial t}\hat{\rho}(t) = [\hat{H},\hat{\rho}(t)]\;.
\end{equation}
% We consider only a real scalar field, for which the Hamiltonian operator can be written as:
% \begin{equation}
% \label{Hamiltonian}
%     \hat{H} = \int  \hat{\mathcal{H}} dx = \int dx [\frac{\hat{\Pi} ^2}{2} + \frac{(\nabla \hat{\phi})^2}{2} + V(\hat{\phi})]\;
% \end{equation}
Combining Eq.~(\ref{W_def}) and Eq.~(\ref{rhoH}), one obtains the equation of motion for the Wigner function~\cite{PhysRevE.49.145,PhysRevLett.123.031601,PhysRevD.50.7542}:
\begin{equation}
    \label{W_eom}
    \frac{\partial}{\partial t} W[\phi ,\Pi;t] = -2H\frac{1}{i \hbar}\sin{(\frac{i \hbar}{2}\Lambda)} W[\phi ,\Pi;t]\;,
\end{equation}
Here, $\Lambda = \frac{\overleftarrow{\delta}}{\delta \Pi} \frac{\overrightarrow{\delta}}{\delta \phi} - \frac{\overleftarrow{\delta}}{\delta \phi} \frac{\overrightarrow{\delta}}{\delta \Pi} $ is the Poisson bracket operator. By expanding the sine function and neglecting terms of order $\hbar^2$ and higher, we recover the classical Liouville equation:
% more details can be found in {\it Supplementary Material }.
% \begin{equation}
%     \label{f2}
%     \begin{split}
%     \frac{\partial}{\partial t} W[\phi ,\pi;t] &= -[H \Lambda - \frac{(i \hbar)^2}{24}H \Lambda^3 + \frac{(i \hbar)^4}{1920}H \Lambda^5 + \cdots ]W[\phi,\Pi ; t] \\
%     &= -(\mathcal{L}_c + \mathcal{L}_q )W[\phi,\Pi ; t]
%     \end{split}
% \end{equation}
% where $\mathcal{L}_c = H \Lambda$ is the classical Liouville operator and $\mathcal{L}_q$ (= all higher-order terms) is the quantum jump operator with $\mathcal{O}(\hbar ^2)$ order precision, using the classical Hamiltonian canonical equation
\begin{equation}
    \label{CL.E}
    \left[ \frac{\partial}{\partial t} + \int dx(\frac{\delta H}{\delta \Pi} \frac{\delta}{\delta \phi} - \frac{\delta H}{\delta \phi} \frac{\delta}{\delta \Pi}) \right] W[\phi,\Pi ; t] = 0\;.
\end{equation}

The classical Liouville equation describes the evolution of the probability distribution function $W[\phi, \Pi ;t]$ over time in classical phase space. This equation can be solved using the method of characteristics, which allows us to determine the evolution of $W$ by following classical trajectories. Under the classical approximation, Eq.~(\ref{CL.E}) can be solved~\cite{JOHN1987152,PhysRevD.50.7542}:
\begin{equation}
W[\phi,\Pi;t]=W[\phi_0,\Pi_0;0],
\label{W_evolve_new}
\end{equation}
where $(\phi_0,\Pi_0)$ denotes the initial phase space point whose classical evolution reaches $(\phi,\Pi)$ after time $t$.
The corresponding classical equations of motion are
\begin{equation}
     \label{HC.E}
     \left\{
     \begin{aligned}
       \frac{\delta H}{\delta \Pi} &= \frac{d \phi}{dt} = \Pi\;, \\
      -\frac{\delta H}{\delta \phi} &= \frac{d\Pi}{dt}=\nabla^2 \phi - \frac{\delta V(\phi)}{\delta \phi} \;,
     \end{aligned}
     \right.
 \end{equation}

Using Eq.~(\ref{W_evolve_new}) and transforming the functional measure along the Hamiltonian flow, one obtains
\begin{equation}
P_{FV}(t)
=
\frac{1}{Z}
\int \mathcal{D}\phi_0 \frac{\mathcal{D}\Pi_0}{2\pi}\,
W[\phi_0,\Pi_0;0]\,
\Theta_{FV}\!\big[\phi_t[\phi_0,\Pi_0]\big].
\label{PFV_initial_measure}
\end{equation}
% % where $J(\phi_{-t},\Pi_{-t})$ is the Jacobian determinant of the measure transformation of the functional integral, which can be defined as
% % \begin{equation}
% %     J(\phi_{-t},\Pi_{-t})=\mathrm{det}
% %     \begin{vmatrix}
% %         \dfrac{\partial \phi}{\partial \phi_{-t}} & \dfrac{\partial \phi}{\partial \Pi_{-t}}\\
% %         \dfrac{\partial \Pi}{\partial \phi_{-t}} & \dfrac{\partial \Pi}{\partial \Pi_{-t}}
% %     \end{vmatrix}
% % \end{equation}
% % In the supplementary material, we prove that $J(\phi_{-t},\Pi_{-t})=1$ through the method of spatiotemporal discretiz ation. In fact, since the phase space flow of the Hamiltonian system is a symplectomorphism and Liuville's theorem ensures the conservation of phase space volume, the Jacobian determinant here is always 1.
Here $\phi_t[\phi_0,\Pi_0]$ denotes the field configuration at time $t$ obtained by evolving the initial condition $(\phi_0,\Pi_0)$ according to Eq.~(\ref{HC.E}). We substitute the result of Eq.~(\ref{W_evolve_new}) and perform a transformation of the integral measure\footnote{The Jacobian of this transformation is unity because Hamiltonian evolution preserves phase space volume, as required by Liouville's theorem.}.
% This can also be verified by evaluating the Jacobian matrix associated with the iterative transformation of the integral measure under the discrete equations of motion.

% \begin{figure}[ht]
%     \centering
%     \includegraphics[width=0.4\textwidth]{figure/schematic_potential.pdf}
%     \caption{The $V$–$\phi$ diagram. $\phi_m$ denotes the field value at the potential maximum; $V_b$ is the barrier height; $\Delta V$ is the potential energy difference between the two vacua; {\it FV} and {\it R} indicate the false vaccum and the true vacuum respectively.}
%     \label{fig:V_phi}
% \end{figure}

%In (1+1) spacetime dimensions, the first-order phase transition proceeds via the nucleation and growth of vacuum bubbles, which in the thin-wall limit take the form of kink-antikink pairs. The decay rate is fully determined by two coupled physical processes: (1) the nucleation of a critical kink-antikink pair via thermal activation or quantum tunneling, which overcomes the free energy barrier to form a metastable critical bubble; (2) the steady expansion of the supercritical bubble, driven by the pressure difference between the false and true vacua, which converts the metastable false vacuum to the stable true vacuum.
%The creation of kink-antikink pairs has been studied with the double-well potential~\cite{Alford:1991qg,Bochkarev:1989tk} and the $\phi^6$ potential~\cite{Dorey:2011yw}. 

In $(1+1)$ spacetime dimensions, a first-order transition proceeds through the nucleation and growth of true-vacuum domains embedded in the false vacuum background. In the thin-wall regime, such a one-dimensional bubble can be viewed as a kink--antikink pair: the kink and antikink form the two domain walls separating the interior true vacuum region from the exterior false vacuum. The real-time conversion of the false vacuum is then governed by two related processes. First, a critical kink--antikink configuration is produced through thermal activation or quantum tunneling, corresponding to a saddle-point configuration that overcomes the free energy barrier. Second, once the separation exceeds the critical size, the resulting supercritical bubble expands under the pressure difference between the false and true vacua, thereby converting the metastable false vacuum into the stable true vacuum. The creation of kink--antikink pairs has been studied in the double well potential~\cite{Alford:1991qg,Bochkarev:1989tk} and the $\phi^6$ potential~\cite{Dorey:2011yw}. 
In this Letter, we study false vacuum decay associated with a first-order transition in the quartic scalar potential: $V(\phi) = \frac{m^2}{2}\phi^2+ \frac{g}{3}\phi^3+ \frac{\lambda}{4}\phi^4 $. The scalar field is assumed to be approximately in thermal equilibrium at the initial time. The density operator is then given by $\hat{\rho} = \frac{1}{Z}e^{-\beta \hat{H}}$, 
% \begin{equation}
%     \label{rho}
%     \hat{\rho} = \frac{1}{Z}e^{-\beta \hat{H}}\;
% \end{equation}
with $\beta=T^{-1}$ in natural units. For the interacting theory, we use the Hartree approximation to approximate the initial thermal density matrix by a self-consistent Gaussian quasi-free state, since a closed analytic expression for the full interacting Wigner functional is not available. We decompose the field into the sum of a uniform background and zero mean fluctuations: $\phi(x)=\varphi+\eta(x), \Pi(x)=\bar{\Pi}+\xi(x)$, with $\langle \eta(x) \rangle = 0, \langle \xi(x) \rangle = 0$, and $G = \langle\eta^2\rangle$ denoting the variance of the field fluctuation. The resulting well-known gap equations are solved iteratively, yielding stable solutions for the Hartree effective mass $M_H$ and the background field $\varphi_\star$.
% \begin{equation}
%     \label{iteration equation}
%     \begin{aligned}
%         &\ G=\langle\eta^2\rangle=\int\frac{dk}{2\pi}\frac{1}{2\omega_k}coth\frac{\omega_k/T}{2}\;\\
%         &\ M_H^2 = m^2 + 2g\varphi + 3\lambda\varphi^2 + 3\lambda G\;\\
%         &\ m^2\varphi+g\varphi^2+\lambda \varphi^3+gG+3\lambda\varphi G=0\
%     \end{aligned}
% \end{equation}
% where $\omega_k=\sqrt{k^2+M_H^2}$ and the solution for the background field is denoted by $\varphi_\star$.
To reduce lattice cutoff dependence in the initial solution, we solve a renormalized gap equation. The numerical results show that our solution has good stability in both the ultraviolet and infrared regions. Detailed information can be found in the {\it Supplementary Material}. Within this Hartree-Gaussian approximation, the initial Wigner functional factorizes into independent momentum modes and is positive definite:
\begin{equation}
W_k(\eta_k,\xi_k)
=
\exp\!\left[
-\tanh\!\left(\frac{\beta\omega_k}{2}\right)
\left(
\frac{|\xi_k|^2}{\omega_k}
+\omega_k|\eta_k|^2
\right)
\right]
\label{Hartree_wigner}
\end{equation}
where $\omega_k=\sqrt{k^2+M_H^2}$. The corresponding momentum space variances of the quantum-thermal distribution are:
\begin{equation}\label{quantum_thermal_corr}
  \begin{aligned}
      &\
      \langle|\eta_k|^2\rangle = \frac{1}{2\omega_k}\coth\frac{\beta \omega_k}{2}  =\frac{1}{\omega_k}\big(\frac{1}{2} + \frac{1}{e^{\omega_k/T}-1}\big)\;, \\
      &\
      \langle|\xi_k|^2\rangle = \frac{\omega_k}{2}\coth\frac{\beta \omega_k}{2}
      = \omega_k\big(\frac{1}{2} + \frac{1}{e^{\omega_k/T}-1}\big)\;.\
   \end{aligned}
\end{equation}
This positive definite initial Wigner functional allows us to sample the initial ensemble directly.
We stress that our numerical procedure only requires positivity of $W[\phi,\Pi;0]$ within the Hartree-Gaussian approximation; it is not necessary to assume that the full time dependent Wigner functional remains an ordinary probability density at all later times.

Numerically, we sample independent Gaussian fluctuations mode by mode according to Eq.~(\ref{quantum_thermal_corr}), perform the inverse Fourier transform to obtain $\eta(x)$ and $\xi(x)$ in real space, and construct the initial fields.
%\begin{equation}
%\phi(x)=\varphi_\star+\eta(x),\qquad \Pi(x)=\bar{\Pi}+\xi(x).
%\end{equation}
Each realization is then evolved with the classical equations of motion, and the false vacuum probability Eq.~(\ref{PFV_initial_measure}) is estimated by the Monte Carlo average:
\begin{equation}
P_{FV}(t)\approx \frac{1}{N_s}\sum_{i=1}^{N_s}
\Theta_{FV}\!\big[\phi_t^{(i)}\big].
\label{num_pfv_new}
\end{equation}
Here, $i$ is used to denote the different Monte Carlo samples of the ensemble, and $N_s$ represents the total number of samples.
The practical construction of the indicator functional $\Theta_{FV}$ is nontrivial because a field configuration is a spatially extended object.
Different operational definitions lead to different observables.
% A survival probability can be built from a sample-level criterion, such as a threshold on a global order parameter or on a connected supercritical cluster.
% Such observables are naturally compared with a Poisson-like exponential decay law over an appropriate intermediate-time interval.
% By contrast, one may also define a fraction-based observable by measuring, for each realization, the fraction of lattice sites that satisfy a local false-vacuum criterion and then averaging over the ensemble.
% This quantity tracks the fraction of space that remains in the false vacuum and is therefore more naturally analyzed using an Avrami-type kinetic law.
In the following sections, we will specify these criteria explicitly and use the corresponding observables as complementary cross-checks of the decay dynamics.

{\it Decay observables and coarse-grained indicators.---}
As discussed above, we use Eq.~(\ref{quantum_thermal_corr}) to perform Gaussian sampling and obtain initial fields containing both quantum and thermal fluctuations. By varying the initial temperature, we probe the decay dynamics in the crossover region between predominantly quantum and predominantly thermal fluctuations. In practice, thermal enhancement is significant mainly for sufficiently low frequency modes, whereas the high $k$ sector remains dominated by the zero-point contribution. Since our lattice implementation regulates the ultraviolet part of the initial variance through the lattice cutoff, the raw pointwise field distribution is not, by itself, a robust decay observable: short wavelength high $k$ modes generate substantial local noise without directly controlling the large scale decay pattern. For this reason, we analyze the decay observables using a coarse-grained field, which suppresses short distance fluctuations and isolates the low energy sector relevant for bubble nucleation, thereby improving the numerical stability and physical interpretability of the simulations.

Specifically, we define the coarse-grained field by
\begin{equation}
\phi_\ell(x,t)=\int dy\,W_\ell(x-y)\,\phi(y,t),
\label{eq:coarse_grained_field}
\end{equation}
where $W_\ell$ is a normalized smoothing kernel, $\int dy\,W_\ell(y)=1$, and $\ell$ denotes the coarse-graining scale. In the numerical implementation, $\ell$ is chosen to be proportional to the false vacuum correlation length,
\begin{equation}
\ell = c_1\,\xi_{\rm fv},
\qquad
\xi_{\rm fv}\sim M_H^{-1},
\label{eq:coarse_scale}
\end{equation}
Unless otherwise stated, the observables entering our decay analysis are constructed from the coarse-grained field rather than from the raw pointwise field. Details of the initial conditions and numerical implementation are given in the {\it Supplementary Material}.

% It is worth emphasizing that this setup should be viewed as a controlled mixed effective description. The low energy background parameters entering the initial ensemble, in particular $M_H$, are obtained from the renormalized Hartree gap equations, whereas the sampled lattice field still retains the lattice regulated zero-point contribution at the level of local pointwise fluctuations. Therefore, local pointwise variances are not themselves the primary physical observables in our analysis. Instead, we focus on coarse-grained fields and on decay observables such as survival probabilities,  transformed spatial fractions, which are much less sensitive to ultraviolet pointwise noise.

We now turn to the implementation of the indicator functional $\Theta_{FV}$. The threshold $\phi_{\rm th}$ on the spatially averaged field has been used as a practical sample-level survival criterion~\cite{PhysRevLett.123.031601,Hertzberg:2020tqa,Pirvu:2023plk,Braden:2017add}, which is useful in regimes where a single bubble dominates the decay in a relatively small box. However, in the general case, especially at higher temperature and for modest vacuum energy difference, the simulation volume may contain multiple bubble seeds. Nearby seeds can collide rapidly, and the post-collision evolution may partially return to the false vacuum. Although the detailed collision dynamics is not the central focus of this work, such local collisions can substantially distort the global spatial average and therefore make a purely whole volume average criterion less reliable.
A more physical sample-level criterion is to monitor whether a sufficiently large true vacuum cluster has emerged and persisted for a sufficiently long time. Accordingly, after coarse graining the field at each time step, we identify connected intervals in which the field exceeds a threshold value $\phi_{\rm th}$. We then introduce a critical length scale $L_c$ and a holding time $\tau_{\rm hold}$. A given realization is declared to have decayed when there first appears a connected interval satisfying:
\begin{equation}
\phi_\ell(x,t)>\phi_{\rm th}
\quad \text{for all } x \text{ in the interval}
\end{equation}
whose length exceeds $L_c$ and whose persistence time exceeds $\tau_{\rm hold}$. In this way, transient short-lived excursions above the threshold are not counted as genuine decay events. The corresponding indicator functional is denoted schematically by $\Theta^{\rm clus}_{FV}\big[\phi_t^{(i)}\big]$: $\Theta^{\rm clus}_{FV}=1$ if no such persistent supercritical cluster has appeared up to time $t$, and $\Theta^{\rm clus}_{FV}=0$ otherwise. Eq.~(\ref{num_pfv_new}) then gives the associated false-vacuum survival probability $P_{\rm surv}(t)$. A related but not identical local box criterion has been discussed in Ref.~\cite{PhysRevD.109.023502}, where the local box-averaged field is used primarily to avoid double counting the same nucleation seed. Our purpose here is different: we use connected supercritical clusters in order to characterize decay in a regime where both single bubble and multi bubble events can contribute.

In a complementary analysis, we additionally implement an independent observable derived from the false vacuum fraction at each time step — an approach well established in prior studies of first-order phase transition dynamics~\cite{PhysRevLett.132.241601,PhysRevE.71.011908,TOMELLINI2022126748}. Unlike the sample-level survival observables from the global and connected-cluster criteria, this observable is preferentially sensitive to the full post-nucleation bubble growth and expansion dynamics, providing a complementary cross-check of the decay dynamics. Specifically, for each coarse-grained sample, we define the false vacuum spatial fraction as:
\begin{equation}
f_{\rm FV}^{(i)}(t)
=
\frac{1}{L}\int dx\,
\Theta\!\big(\phi_{\rm th}-\phi_\ell^{(i)}(x,t)\big)
\label{eq:false_fraction}
\end{equation}
namely the fraction of space that lies on the false vacuum side of the threshold. Averaging over the ensemble yields:
\begin{equation}
P_{\rm frac}(t)
=
\frac{1}{N_s}\sum_{i=1}^{N_s} f_{\rm FV}^{(i)}(t)
\label{eq:pfrac_def}
\end{equation}

% Unlike the survival observables, $P_{\rm frac}$ is not a sample-level survival probability; it measures the ensemble averaged false vacuum spatial fraction and is therefore fitted with an Avrami type law.

{\it Numerical simulation.---}
 Unlike the analog potentials proposed to simulate vacuum decay in cold atom experiments~\cite{Braden:2017add,Billam:2020xna,Billam:2018pvp,Billam:2021qwt}, here we consider a scalar potential that is conventionally used to study gravitational waves from first-order phase transitions in the early Universe~\cite{Cutting:2018tjt,Cutting:2020nla,Bian:2025twi}.
All real-time FVD lattice simulations are performed on a (1+1)D lattice in natural units. We adopt a dimensionless rescaling scheme to simplify numerical implementation: $\tilde{t} = m t$, $\tilde{x} = mx$ and $\tilde{T}=T/m$, $\tilde{\phi} = \phi/\phi_0$, $\tilde{\Pi}=\frac{\Pi}{m\phi_0}$, where the scalar field is dimensionless, and we fix the scaling factor $\phi_0=1$ for simplicity. 
This rescaling fully eliminates all dimensionful parameters from the equations of motion, significantly improving the stability and efficiency of our lattice simulations. The Lagrangian density reads:
\begin{equation}
\label{Lagrangian}
\begin{split}
    \mathcal{L}&=\frac{1}{2} \partial^\mu\phi\partial_\mu\phi-(\frac{m^2}{2}\phi^2+\frac{g}{3}\phi^3+\frac{\lambda}{4}\phi^4)\\ &= m^2[\frac{1}{2}\dot{\tilde{\phi}}^2-\frac{1}{2}(\nabla \tilde{\phi})^2-(\frac{1}{2}\tilde{\phi}^2+\frac{\tilde{g}}{3}\tilde{\phi}^3+\frac{\tilde{\lambda}}{4}\tilde{\phi}^4) ]\;
\end{split}
\end{equation}
% Here we take $\tilde{g}=g/m^2=-0.5888, \tilde{\lambda}=\lambda/m^2=0.0703$.
We fix the dimensionless coupling constants as \(\tilde{g} = g/m^2 = -0.5888\) and \(\tilde{\lambda} = \lambda/m^2 = 0.0703\). For brevity, we omit the tilde accent on all dimensionless quantities in what follows. 
We set the dimensionless integration time step to $\Delta t = 0.01$, the lattice spacing to $a = 0.25$, and the number of lattice points to $N_x = 1024$.
To ensure robust statistical convergence of our results, we generate \(N_s = 10^5\) statistically independent initial  $(\phi,\Pi)$ configurations. For each individual configuration, we solve the following equation of motion using the fourth-order Runge-Kutta method, which we employ to evolve the field configuration $\phi_t^{(i)}$
that enters the indicator functional $\Theta_{FV}$:
\begin{equation}
    \label{EOM_specific}
     \begin{aligned}
       \frac{\partial^2 \phi}{\partial t^2} &= \frac{\partial^2\phi }{\partial x^2} - (\phi+g\phi^2+\lambda\phi^3)
     \end{aligned}
\end{equation}

After obtaining the real-time evolved field configurations, we coarse-grain the raw field via Eq.~(\ref{eq:coarse_grained_field}). We then adopt three distinct criteria with a threshold set as \(\phi_{\rm th} = \phi_b + c_2\), where \(\phi_b\) denotes the field value at the potential barrier. For the connected-cluster criterion, we take the critical length scale \(L_c = c_3 \xi_{\rm fv}\) and the holding time \(\tau_{\rm hold} = c_4 \xi_{\rm fv}\).
This procedure yields three distinct real-time observables, namely \(P^{\rm glob}_{\rm surv}(t)\), \(P^{\rm clus}_{\rm surv}(t)\), and \(P_{\rm frac}(t)\).
The three distinct implementations of the false vacuum indicator 
\(\Theta_{\rm FV}\)
 lead to qualitatively different temporal evolution of the false vacuum decay signal, rooted in the fundamentally distinct physical meaning of the corresponding observables. Specifically, the global-average and connected-cluster criteria define binary sample-level survival observables, which classify each full lattice configuration as either decayed or remaining in the false vacuum, and quantify the fraction of surviving samples. By contrast, the fraction-based approach is a spatial observable, measuring the volume fraction of the lattice that remains on the false vacuum side of the potential barrier at each time step.
 
%After obtaining the real-time evolving field, we use Eq.~(\ref{eq:coarse_grained_field}) to obtain the coarsened field, and then by using three different criteria, the threshold we set is $\phi_{th} = \phi_b + c_2$, $\phi_b$ represents the field value at the barrier. For the scheme of the cluster criterion, the critical length we take is $L_c = c_3  \xi_{\rm fv}$, and the holding time is $\tau_{\rm hold} = c_4  \xi_{\rm fv}$. We can obtain three different $P_{FV}(t)$, namely $P^{\rm glob}_{\rm surv}(t),~ P^{\rm clus}_{\rm surv}(t),~P_{\rm frac}(t)$, the three implementations of $\Theta_{FV}$ lead to qualitatively different time dependences of the false vacuum signal. The global-average criterion and the connected-cluster criterion define sample-level survival observables, while the fraction-based definition measures the fraction of space that remains on the false vacuum side.
For the sample-level survival observables, one can extract the decay rate per unit length from $\ln P_{\rm surv}(t)=-\Gamma_{\rm surv}\,L\,t+C,$ 
within a suitable intermediate time window ~\cite{PhysRevLett.123.031601,Hertzberg:2020tqa,Pirvu:2023plk}, where $L$ is the lattice size and $C$ is a constant. 
This method allows one to define an effective direct decay rate $\Gamma$, as emphasized in Refs.~\cite{PhysRevLett.117.231601,PhysRevD.95.085011}. For survival observables, we fit the time intervals in which $\ln P_{\rm surv}^{\rm glob}(t)$ and $\ln P_{\rm surv}^{\rm clus}(t)$ are approximately linear in $t$, excluding the early transient from short-lived clusters and the late tail with poor survival statistics. 
The fitting uncertainty is estimated by varying the subwindow inside the admissible range. We also varied the coarse-graining scale, threshold offset, critical cluster length, and holding time. The extracted temperature dependence remains stable within the explored range. Increasing the connected-cluster length scale $L_c$ delays the cluster decay assignment and makes the criterion more restrictive, but its relation to the global-survival result is temperature dependent: in the multi-bubble regime, larger $L_c$ can move the cluster curve toward, or even slower than, the global criterion. Representative parameter scans and fitting details are shown in the {\it Supplementary Material}.
For the false vacuum fraction based observable, we instead use the form of the Kolmogorov–Johnson–Mehl–Avrami (KJMA) formula in (1+1)D  ~\cite{Witten:1984rs,Turner:1992tz}: $\ln P_{\rm frac}(t)=-\Gamma_{\rm frac}\,v\,t^2+C$ within a suitable intermediate time window avoiding late-time contamination from bubble collisions, oscillations, and partial returns toward the false vacuum side, where $v$ is the bubble wall velocity estimated from the time evolution of several randomly selected coarse-grained sample profiles by tracking the growth of supercritical domains. The $P_{\rm frac}(t)$ counts local conversion immediately through the false vacuum spatial fraction, while the global-average survival changes only when the converted fraction is large enough to shift the sample-averaged field across the threshold. Thus $\Gamma_{\rm frac}$ should be interpreted as a KJMA kinetic measure of spatial conversion and vacuum domain growth, complementary to the sample-level survival rates.

To compare our simulation results with the finite temperature theoretical predictions from thermal equilibrium field theory, i.e., the Langer-Affleck rate~\cite{Affleck:1980ac,Langer:1969bc,Linde:1981zj}, we compute the leading semiclassical decay rate by obtaining the instanton solution following Refs.~\cite{Ekstedt:2023sqc,Wang:2026ifp}: $\Gamma = A_{\rm stat}A_{\rm dyn}e^{-S/T}\;$. Here, $A_{\rm stat}$ is the statistical factor from the fluctuation determinant, and the dynamical factor $A_{\rm dyn}=\frac{1}{2\pi}(\sqrt{\vert\lambda_{-}\vert+\eta^2/4}-\eta/2)$ with
$\lambda_{-}$ being the negative eigenvalue of the functional determinant, the Langevin damping coefficient $\eta$ should be determined by the real-time simulation. The bounce action $S$ in $(1+1)$ dimensional spacetime reads
\begin{equation}
    \label{SB}
    S=\int dr\left[\frac{1}{2}(\frac{\partial \phi_B}{\partial r})^2+V_{eff}(\phi_B)\right]\;,
\end{equation}
where $\phi_B$ is the bounce solution of
\begin{equation}
    \label{bounce}
    \frac{\partial^2 \phi_B}{\partial r^2}-\frac{\partial V_{eff}}{\partial\phi}=0\;.
\end{equation}
where \(V_{\rm eff}\) is the Hartree effective potential; further details are given in the {\it Supplementary Material}.
Here, the temperature should be interpreted as the effective temperature in our simulation. Since the initial ensemble contains both thermal and quantum fluctuations, the input temperature $T_{\rm ini}$ is not the most direct measure of the thermal energy stored in the long-wavelength modes that control the nucleation events. We therefore introduce an effective temperature $T_{\rm eff}$ as a diagnostic of the low-k modes. Concretely, we monitor the ensemble-averaged momentum spectrum during the early-time evolution and extract $T_{\rm eff}$ from the low momentum window in which the spectrum approximately follows the equipartition theorem: ${\cal P}_{\Pi}(k,t)\simeq T_{\rm eff}(t),\quad k<k_{\rm cut}$. The cutoff $k_{\rm cut}$ is chosen to be of order the inverse critical bubble scale, with $R_c\simeq 2.44$ in the present potential. We then identify an early time plateau of $T_{\rm eff}(t)$ before substantial false vacuum decay occurs and use this value as the effective temperature. The detailed extraction procedure and comparisons among alternative definitions of $T_{\rm eff}$ are given in the {\it Supplementary Material}.

\begin{figure}[!htp]
    \centering
    \includegraphics[width=0.46\textwidth]{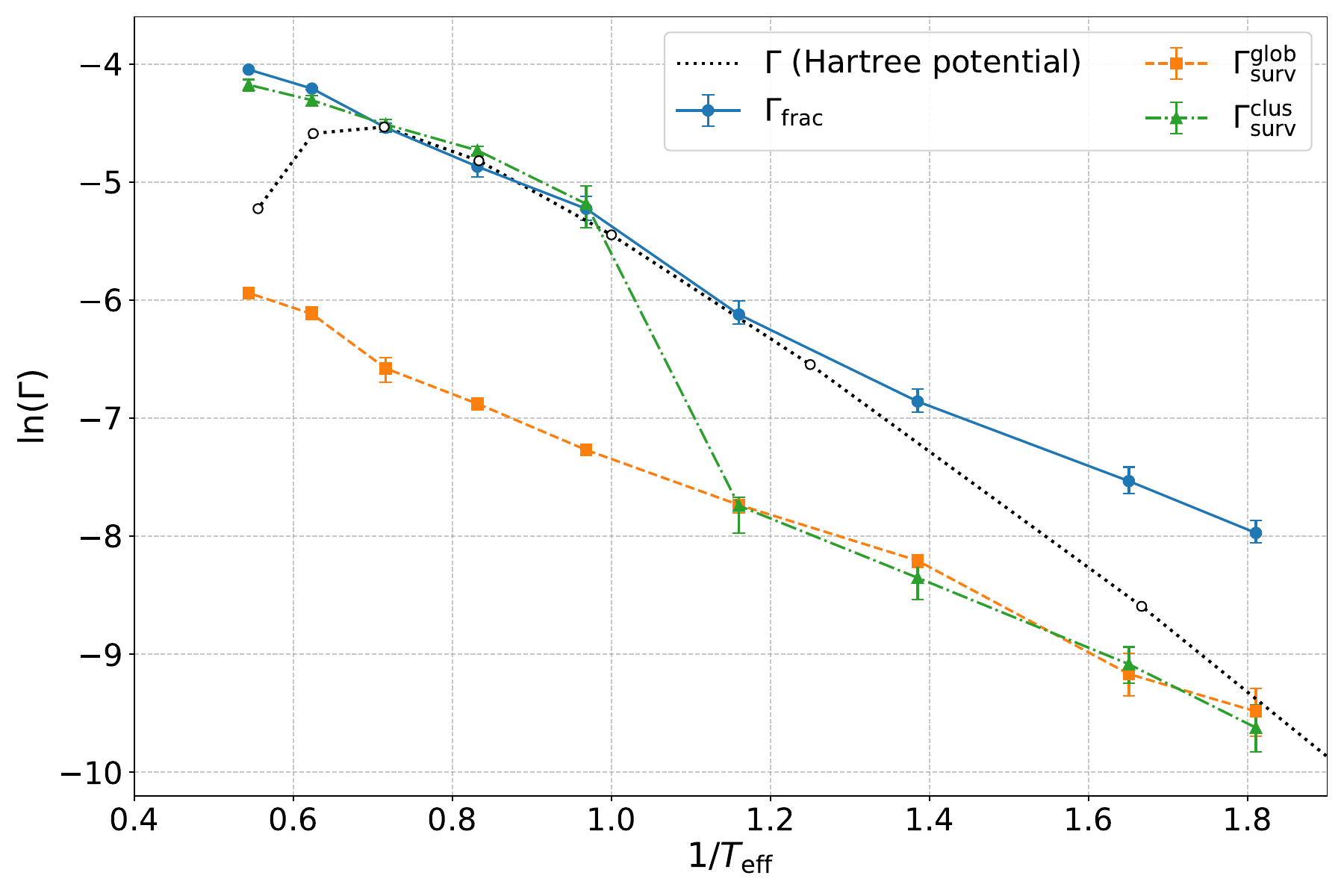}
    \caption{Comparison between simulation results and theoretical predictions with $\eta=0$. The quoted uncertainty is taken as the larger of the regression error and the fit-window systematic obtained from subwindow refits within the selected fitting range, see the fit of the $P_{frac},P_{\rm surv}^{\rm glob},P_{\rm surv}^{\rm clus}$ in {\it Supplemental Material} for details. 
    % Here we take $c_1=3,~c_2=0.8,~c_3=10,~c_4=3$.
    }
    \label{fig:gamma_t}
\end{figure}

Fig.~\ref{fig:gamma_t} shows the FVD rates extracted from three different methods as a function of $1/T_{\rm eff}$, compared with the theoretical predictions from the Langer-Affleck formalism using the Hartree effective potential. In the high-temperature thermal nucleation regime ($1/T_{\rm eff} < 1$), the KJMA description is a good effective description over the fitted intermediate time window, and the decay rates extracted from the fraction method and connected-cluster method are broadly consistent with the Hartree thermal benchmark, supporting the validity of our numerical approach and the Hartree-Gaussian initial ensemble. Here, we note that the semiclassical thermal-nucleation construction becomes less stable when the Hartree effective barrier becomes shallow at high temperature. This can lead to the turnover of the Hartree benchmark. However, in this regime, the global-average criterion introduces a spurious systematic delay in the assignment of decay events: by spatially averaging over regions that have already decayed to the true vacuum and those remaining in the false vacuum, it fails to capture the onset of nucleation until the majority of the lattice has transitioned.
This qualitative difference in decay dynamics explains the systematic discrepancy between the global survival method and our connected-cluster method and fraction method at high temperatures.

As the temperature decreases and the system enters the quantum tunneling regime, significant discrepancies emerge between the different methods. The false-vacuum fraction observable can overestimate the sample-level decay rate, especially at the lowest temperatures studied.
This behavior is associated with short-lived subcritical kink--antikink-like domains and kink--antikink reflection or partial-return dynamics, which can move transformed regions back toward the false-vacuum side and thereby violate the irreversibility assumption underlying the KJMA interpretation.
In contrast, the sample-level survival methods are less sensitive to these transient spatial-conversion effects. When nucleation events are sufficiently dilute, the global and connected-cluster criteria identify nearly the same decay events, with only a small difference in the assigned decay time.

The Hartree-potential prediction should be regarded as a thermal benchmark rather than an exact result. It is based on a spatially homogeneous, Gaussian self-consistent background and therefore does not include all fluctuation-determinant, dynamical-prefactor, and real-time non-equilibrium effects. In particular, it does not describe the detailed kink--antikink reflection or partial-return dynamics visible in the real-time simulations. For this reason, an overall normalization uncertainty remains in the theoretical curve. In Fig.~\ref{fig:gamma_t}, we include a constant normalization factor when comparing the Hartree benchmark with the simulation data. This factor affects the overall scale but not the temperature dependence of the benchmark. The remaining discrepancies, especially at low temperatures, should therefore be interpreted as reflecting both the limitations of the benchmark and the observable dependence of the real-time rate extraction.

Additional uncertainties may arise because the metastable phase need not remain in full thermal equilibrium during real-time evolution~\cite{Pirvu:2024nbe,Pirvu:2024ova,Hirvonen:2025hqn}\footnote{ A key methodological difference from Ref.~\cite{Hirvonen:2025hqn} is that we employ a Wigner-functional approach that includes zero-point fluctuations in the initial ensemble, whereas that work used a classical Monte Carlo sampling of the $\exp(-\beta H)$ distribution, which is valid only in the high-temperature classical regime where quantum zero-point fluctuations are negligible. Furthermore, we employ a non-symmetric double-well potential featuring a metastable false vacuum, which exhibits a first-order phase transition, whereas Ref.~\cite{Hirvonen:2025hqn} considered a symmetric double-well potential.}. Oscillon-like configurations that can arise from quantum fluctuations prior to bubble nucleation may contribute to residual systematic uncertainties in the extracted decay rate~\cite{Bian:2025twi,Gleiser:2007ts}, resulting in the discrepancy between the numerically measured and the theoretical prediction of the FVD rate.

{\it Conclusions and discussions.---}
In this Letter, we developed a real-time Wigner functional lattice approach to compute the false vacuum decay rate at finite temperature. Within the Hartree approximation, the interacting initial state is represented by a self-consistent Gaussian Wigner ensemble, allowing the false vacuum basin probabilities to be evaluated by Monte Carlo sampling. We propose the connected-cluster survival method, which provides a physically interpretable numerical framework for studying false vacuum decay across the temperature range studied. We find both the false vacuum fraction method and the connected-cluster survival criterion achieve excellent quantitative agreement with the Langer rate in the thermal nucleation regime, and the time delay between bubble nucleation and global field crossing leads to a significant underestimation of the true instantaneous nucleation rate by the global survival method in the regime. At lower temperatures, the connected-cluster criterion yields results consistent with the global-survival method, and subcritical or marginal kink--antikink-like domains can collapse, oscillate, or partially return to the false vacuum side invalidates the false vacuum fraction method. While the coherent quantum reflection effect is unique to 1+1D topological kink-antikink collisions, the spurious signal from subcritical fluctuations is a general phenomenon that occurs in any dimension. Therefore, similar transient subcritical fluctuations may also affect fraction-based diagnostics in higher dimensions, although the magnitude and detailed mechanism may differ.

Our connected-cluster survival criterion provides a real-time, local diagnostic for identifying persistent supercritical domains and extracting false vacuum decay rates in regimes where global survival can be affected by multi-bubble dynamics. This construction is not restricted to one spatial dimension: in higher-dimensional scalar theories, the connected interval can be replaced by connected supercritical regions or volumes, making the method relevant to numerical studies of cosmological first-order phase transitions and metastable-vacuum decay~\cite{Coleman:1977py,Callan:1977pt,Langer:1969bc,Affleck:1980mp}. It can therefore complement existing saddle-point and real-time approaches, especially when inhomogeneous field configurations, bubble growth, and domain conversion are important~\cite{Andreassen:2016cff,Braden:2018tky}. Possible applications include real-time estimates of nucleation and conversion dynamics entering gravitational-wave predictions from cosmological phase transitions~\cite{Kosowsky:1991ua,Kamionkowski:1993fg,Caprini:2015zlo,Caprini:2019egz}, electroweak baryogenesis scenarios involving expanding bubble walls~\cite{Rubakov:1996vz,Morrissey:2012db}, and first-order transitions that may seed primordial black holes~\cite{Carr:2020xqk,Gouttenoire:2023naa,Liu:2021svg,Lewicki:2023ioy}. More generally, the same connected-domain logic can be adapted to systems where decay is diagnosed by the emergence and persistence of supercritical bubbles, defects, or ordered domains~\cite{Zurek:1996sj,Anglin:1998pm}.

%Our initial configurations for the scalar field and its conjugate momentum are similar to those of Ref.~\cite{Hirvonen:2025hqn}. Our results show that the global-survival method yields results similar to those of Refs.~\cite{Alford:1991qg,Bochkarev:1989tk} adopt the Langevin technique: the false vacuum decay rate is much smaller than the theoretical prediction. By contrast, in the high-temperature regime, both the false vacuum fraction method and the connected-cluster survival method agree consistently with the thermal nucleation theory except that of an extra suppression factor .
%Oscillon-like configurations that appear before bubble nucleation may further affect the extracted rate~\cite{Bian:2025twi,Gleiser:2007ts}, which can lead to a discrepancy between the prefactor extracted in our simulation and the analytical prediction.
%Future extensions to 3+1 dimensions could be applied to gravitational-wave production from cosmological phase transitions and to baryon number violating processes such as the sphaleron rate~\cite{Khlebnikov:1988sr}.

{\it Acknowledgments.---}
This work is supported by the National Natural Science Foundation of China (NSFC) under Grants Nos.  12322505, and 12347101.
L.B. also acknowledges Chongqing Natural Science Foundation under Grant
No. CSTB2024NSCQ-JQX0022 and
Chongqing Talents: Exceptional Young Talents Project No. cstc2024ycjh-bgzxm0020.

\bibliography{references}

\clearpage

\appendix
\begin{onecolumngrid} % 附录单栏模式
\section{Supplementary Material}

This Supplemental Material provides technical details supporting the main text, including the Hartree-Gaussian initial state, renormalized lattice implementation, coarse-grained observables, thermal benchmark calculation, fitting prescriptions, parameter dependence checks, and effective temperature extraction.

\section{Hartree-Gaussian initial state and renormalized lattice implementation}
\label{app:hartree_init}

For the interacting scalar theory, the exact thermal Wigner functional associated with
$\hat{\rho}=\frac{1}{Z}e^{-\beta \hat H}$ cannot be written in closed form. In order to construct a tractable initial ensemble, we adopt a Hartree-Gaussian approximation, i.e. a self-consistent quasi-free approximation to the interacting thermal state. This yields a positive-definite initial Wigner functional and allows direct mode-by-mode Monte Carlo sampling. The logic is closely related to the Hartree treatment of self-consistent propagators and effective masses \cite{Cornwall:1974vz,BERGES2005344,Reinosa:2011ut}.

\paragraph{Field decomposition and Gaussian factorization.}
We decompose and expand the original potential into uniform background field $\varphi$ and fluctuation field $\eta(x)$ up to the fourth order:
\begin{equation}
\begin{aligned}
V(\varphi+\eta)
&=
V(\varphi)+V'(\varphi)\eta+\frac12V''(\varphi)\eta^2
+\left(\frac{g}{3}+\lambda\varphi\right)\eta^3+\frac{\lambda}{4}\eta^4 .
\end{aligned}
\end{equation}
For the thermal initial state considered here, we set $\bar\Pi=0$. Within the Hartree-Gaussian approximation, higher fluctuation moments are factorized using Wick contraction:
\begin{equation}
\eta^3 \rightarrow 3\langle \eta^2\rangle \eta = 3G\,\eta,
\qquad
\eta^4 \rightarrow 6\langle \eta^2\rangle \eta^2 - 3\langle \eta^2\rangle^2
=6G\,\eta^2-3G^2,
\label{eq:wick_hartree_rules}
\end{equation}
which is the local Gaussian closure underlying the Hartree truncation. One then obtains
\begin{equation}
V(\varphi+\eta)\approx U_H(\varphi,G)+A(\varphi,G)\,\eta+\frac12 M_H^2(\varphi,G)\,\eta^2,
\end{equation}
with
\begin{equation}
    A(\varphi,G)=m^2\varphi+g\varphi^2+\lambda\varphi^3+g G+3\lambda\varphi G
\end{equation}
\begin{equation}
M_H^2(\varphi,G)=m^2+2g\varphi+3\lambda\varphi^2+3\lambda G.
\label{eq:MH_def}
\end{equation}
% and $U_H(\varphi,G)=V(\varphi)-\frac{3}{4}\lambda G^2$ collects all terms independent of the fluctuation field after the Hartree truncation.
The term $U_H(\varphi,G)$ collects all terms independent of the fluctuation field after the Hartree truncation.
Its explicit form depends on whether one keeps the fluctuation bilinear $\eta^2$ as an operator term or partially averages it into the background free-energy density.
For the present construction, only $A(\varphi,G)$ and $M_H^2(\varphi,G)$ are needed to determine the self-consistent background and the Gaussian fluctuation mass.

\paragraph{Tadpole condition and self-consistent Hartree mass.}
The background field must be chosen such that the fluctuation has vanishing one-point function. This is equivalent to requiring that the coefficient of the linear term in $\eta$ vanish: $A(\varphi,G)=0$, namely:
\begin{equation}
m^2\varphi+g\varphi^2+\lambda\varphi^3+gG+3\lambda\varphi G=0.
\label{eq:tadpole_condition}
\end{equation}

\begin{figure}[htbp]
\centering
\begin{minipage}{0.45\textwidth}
\centering
\begin{tikzpicture}[scale=1.0]
    % external line
    \draw[dashed, thick] (-2,0) -- (0,0);

    % tadpole loop attached at the endpoint
    \draw[dashed, thick] (0,0) .. controls (0.55,0.65) and (0.55,1.35) .. (0,1.6)
                        .. controls (-0.55,1.35) and (-0.55,0.65) .. (0,0);

    % vertex
    \filldraw (0,0) circle (1.3pt);

    % labels
    \node at (-0.8,-0.32) {$\varphi$};
    \node at (0,1.95) {$G$};
\end{tikzpicture}
\end{minipage}
\hfill
\begin{minipage}{0.45\textwidth}
\centering
\begin{tikzpicture}[scale=1.0]
    % external line
    \draw[dashed, thick] (-2,0) -- (0,0) -- (2,0);

    % tadpole loop attached at the center
    \draw[dashed, thick] (0,0) .. controls (0.55,0.65) and (0.55,1.35) .. (0,1.6)
                        .. controls (-0.55,1.35) and (-0.55,0.65) .. (0,0);

    % vertex
    \filldraw (0,0) circle (1.3pt);

    % labels
    \node at (-0.7,-0.32) {$\eta$};
    \node at (0.7,-0.32) {$\eta$};
    \node at (0,1.95) {$G$};
\end{tikzpicture}
\end{minipage}
\caption{Tadpole diagrams entering the Hartree approximation. Left: one-point tadpole contribution to the self-consistent background equation. Right: two-point tadpole self-energy contribution to the fluctuation mass.}
\label{fig:tadpole_diagrams}
\end{figure}

This condition guarantees that $\varphi$ is a self-consistent mean field. Diagrammatically, Eq.~(\ref{eq:tadpole_condition}) states that the sum of one-point tadpole insertions vanishes after the background has been shifted to $\varphi=\varphi_\star$, see the left of Fig.~\ref{fig:tadpole_diagrams}. Likewise, the corresponding two-point tadpole insertion generates the local Hartree self-energy shift $3\lambda G$ entering Eq.~(\ref{eq:MH_def}), see the right of Fig.~\ref{fig:tadpole_diagrams}.
Once the fluctuation Hamiltonian has been reduced to quadratic form, each momentum mode behaves as an independent harmonic oscillator with frequency
\begin{equation}
\omega_k=\sqrt{k^2+M_H^2}.
\end{equation}
The variance of the fluctuation field in the thermal Gaussian state is therefore
\begin{equation}
G=\langle \eta^2\rangle
=\int\frac{dk}{2\pi}\frac{1}{2\omega_k}\coth\frac{\beta\omega_k}{2}.
\label{eq:G_cont}
\end{equation}
Equations~(\ref{eq:MH_def}), (\ref{eq:tadpole_condition}), and (\ref{eq:G_cont}) form the closed Hartree system quoted in the main text. Their solution determines the self-consistent background $\varphi_\star$, the fluctuation variance $G$, and the Hartree effective mass $M_H$.

\paragraph{Renormalized lattice gap equation.}
On the lattice, with spacing $dx$ and total length $L=N_x\,dx$, the momentum and lattice dispersion are
\begin{equation}
k_n=\frac{2\pi n}{L},
\qquad
\hat k_n^2=\frac{4}{dx^2}\sin^2\frac{k_n dx}{2},
\end{equation}
so that
\begin{equation}
\omega_n=\sqrt{\hat k_n^2+M_H^2}.
\end{equation}
The naive lattice variance,
\begin{equation}
G_{\rm bare}(M_H,T)=\frac{1}{L}\sum_n\frac{1}{2\omega_n}\coth\frac{\beta\omega_n}{2},
\label{eq:G_bare}
\end{equation}
contains the ultraviolet zero-point contribution and induces an unphysical drift of the self-consistent solution with the lattice cutoff. To suppress this cutoff sensitivity, we solve instead a subtracted/renormalized gap equation,
\begin{equation}
G_{\rm sub}(M_H,T)=\frac{1}{L}\sum_n
\left[
\frac{1}{2\omega_n}\coth\frac{\beta\omega_n}{2}
-\frac{1}{2\omega_{n,{\rm ref}}}
\right],
\label{eq:G_sub}
\end{equation}
where
\begin{equation}
\omega_{n,{\rm ref}}=\sqrt{\hat k_n^2+m_{\rm ref}^2}.
\end{equation}
In practice, $G_{\rm sub}$ replaces $G$ in the self-consistent iteration. For the reference mass $m_{\rm ref}$, the curvature near the tree-level false vacuum that we take is: $m^2_{\rm ref} = 1$. The subtraction removes the reference-vacuum ultraviolet drift from the low-energy Hartree background parameters while preserving the thermal contribution. Numerically, the coupled equations are solved by fixed-point iteration with under-relaxation until $\varphi_\star$, $G$, and $M_H^2$ converge.

To verify that the renormalized Hartree construction is numerically stable, we study the dependence of $\varphi_\star$ and $M_H$ on both the ultraviolet cutoff (at fixed physical volume) and the infrared box size (at fixed lattice spacing). We find that the renormalized solution exhibits good convergence in both limits, as shown in Fig.~\ref{fig:hartree verification}:
\begin{figure}[H]
    \centering
    \includegraphics[width=0.8\linewidth]{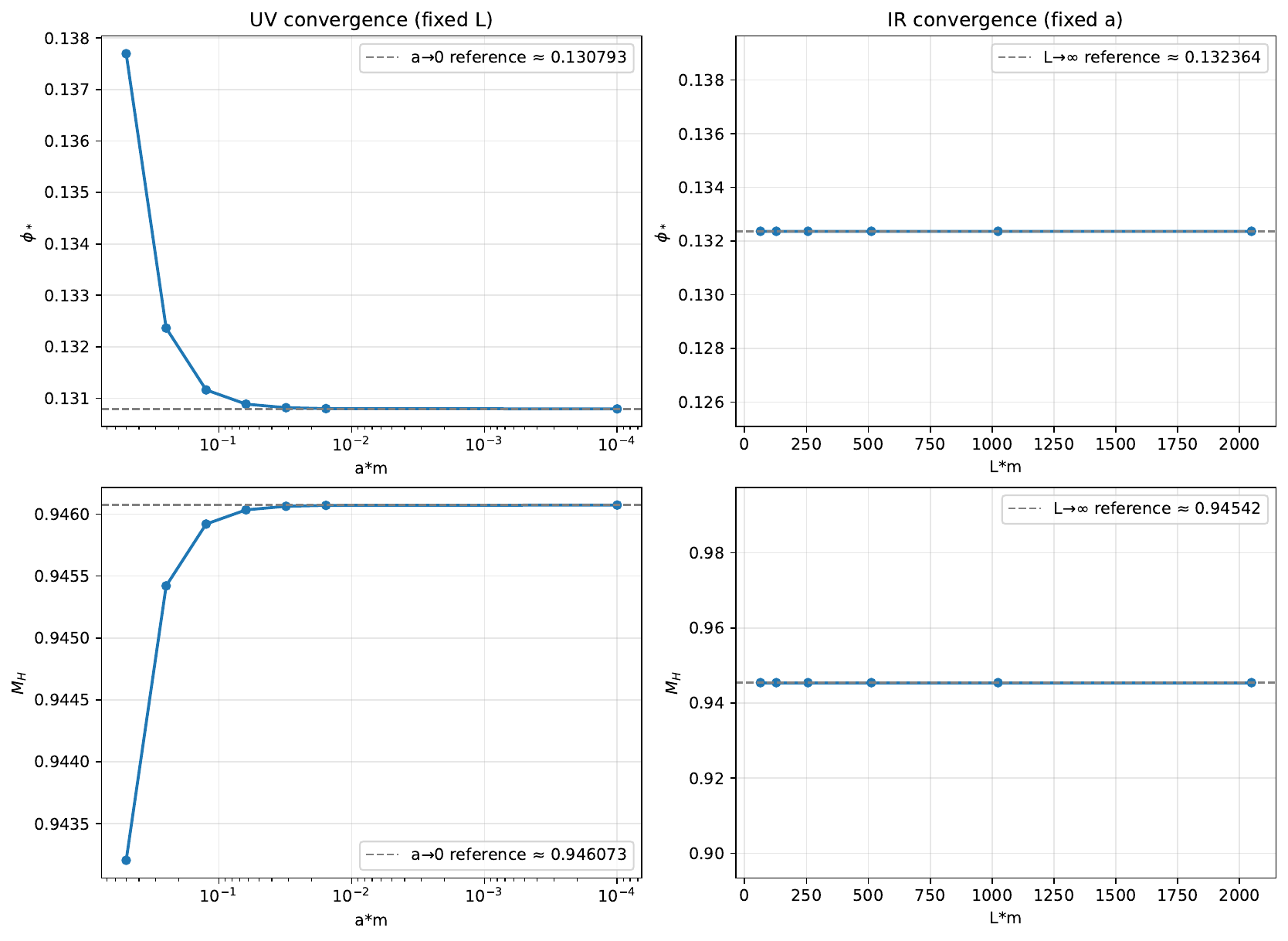}
    \caption{Verification of ultraviolet/infrared convergence after lattice subtraction. We use $T/m=1$ as an example, in UV and IR convergence, we fixed $L*m=256, a*m=0.25$ respectively.}
    \label{fig:hartree verification}
\end{figure}

This confirms that the initial ensemble used in the simulations is controlled at the level of the low-energy Hartree background parameters, even though local pointwise field variances remain those of a lattice-regulated Gaussian random field.
% In the present implementation, the subtraction is imposed at the level of the self-consistent Hartree background and effective mass. The sampled lattice field remains a lattice-regulated Gaussian random field, so local pointwise variances are not themselves cutoff-independent observables. The physically relevant quantities used in the analysis are the renormalized Hartree background parameters and coarse-grained observables.
Within the Hartree-Gaussian approximation, the initial density operator factorizes into independent momentum modes, giving the mode variances quoted in Eq.~(10) of the main text. We sample independent Gaussian modes according to these variances, impose the reality conditions $\eta_{-k}=\eta_k^\ast$ and $\xi_{-k}=\xi_k^\ast$, and then perform the inverse Fourier transform to obtain the real-space initial fields.
% The subsequent real-time evolution is then performed with the classical equations of motion of the original interacting Hamiltonian, while the initial ensemble is entirely specified by the renormalized Hartree-Gaussian approximation.
% In the present implementation, the subtraction is imposed at the level of the self-consistent Hartree background and effective mass. The sampled lattice field remains a lattice-regulated Gaussian random field, so local pointwise variances are not themselves cutoff-independent observables. The physically relevant quantities used in the analysis are the renormalized Hartree background parameters and coarse-grained observables.

\section{Implementation of coarse graining and initial field diagnostics}
\label{app:coarse_graining}

In the main text, the decay observables are constructed from a coarse-grained field in order to suppress ultraviolet pointwise noise and isolate the long wavelength structures relevant for false vacuum decay. Here we describe the numerical implementation of the coarse-graining procedure and present several diagnostics of the initial ensemble and subsequent real-time evolution.

In the continuum notation used in the main text, the coarse-grained field is defined in Eq.~(13). In the actual numerical implementation, we use box smoothing on the lattice. For a lattice field $\phi_i(t)\equiv \phi(x_i,t)$, the coarse-grained field is defined by a local moving average:
\begin{equation}
\phi^{(\ell)}_i(t)
=
\frac{1}{N_\ell}
\sum_{m=-M}^{M}
\phi_{i+m}(t),
\label{eq:cg_box_discrete}
\end{equation}

where periodic boundary conditions are imposed on the index $i+m$, and $N_\ell=2M+1$ is the number of lattice points inside the coarse-graining window. The physical coarse-graining scale is therefore $\ell\simeq N_\ell a$, with $a$ the lattice spacing. In the simulations, $\ell$ is chosen to be proportional to the false vacuum correlation length, as described in the main text.
We stress that the coarse-grained field is used as an analysis field, not as a new dynamical variable. The microscopic field $\phi(x,t)$ is always evolved with the original equations of motion. Coarse graining is applied only when constructing observables, such as the connected-cluster survival criterion or the fraction-based false vacuum indicator. In this sense, the coarse-graining scale $c_1$ should be interpreted as a resolution parameter for the observables rather than as an additional dynamical coupling.

Figure~\ref{fig:init_phi} shows the initial field value distributions constructed from the coarse-grained field for several temperatures and for three representative choices of the coarse-graining coefficient, $c_1=2,3,4$. In all cases, the distribution broadens as the temperature increases, reflecting the increasing importance of thermal fluctuations. At fixed temperature, the distribution becomes visibly narrower as $c_1$ increases. This behavior is expected: a larger coarse-graining window averages over a larger neighborhood and therefore suppresses more short wavelength fluctuations. Equivalently, increasing $c_1$ lowers the effective spatial resolution of the observable, so that the coarse-grained field retains only the smoother, long wavelength content of the configuration.

\begin{figure}[H]
    \centering
    \includegraphics[width=0.3\linewidth]{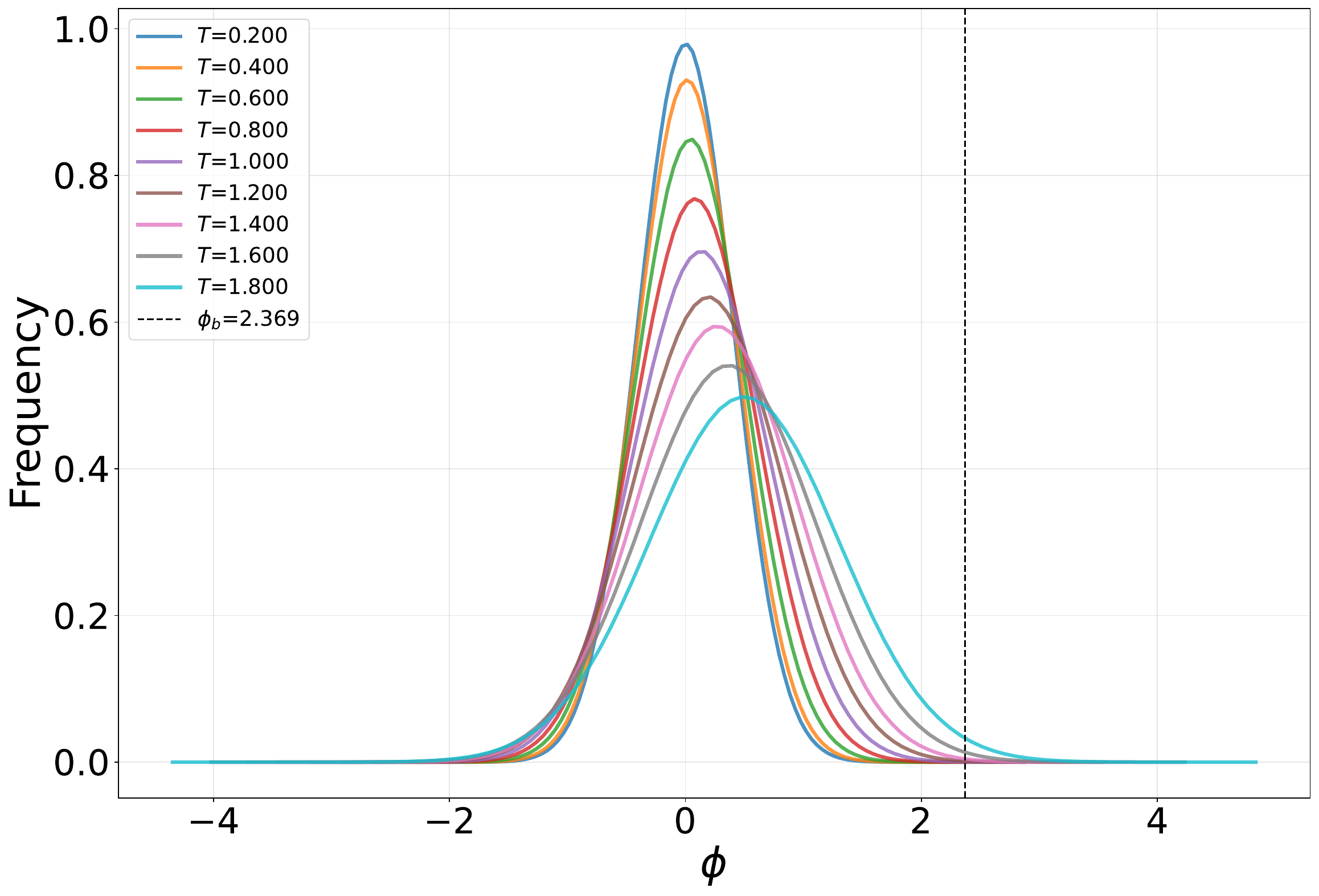}
    \includegraphics[width=0.3\linewidth]{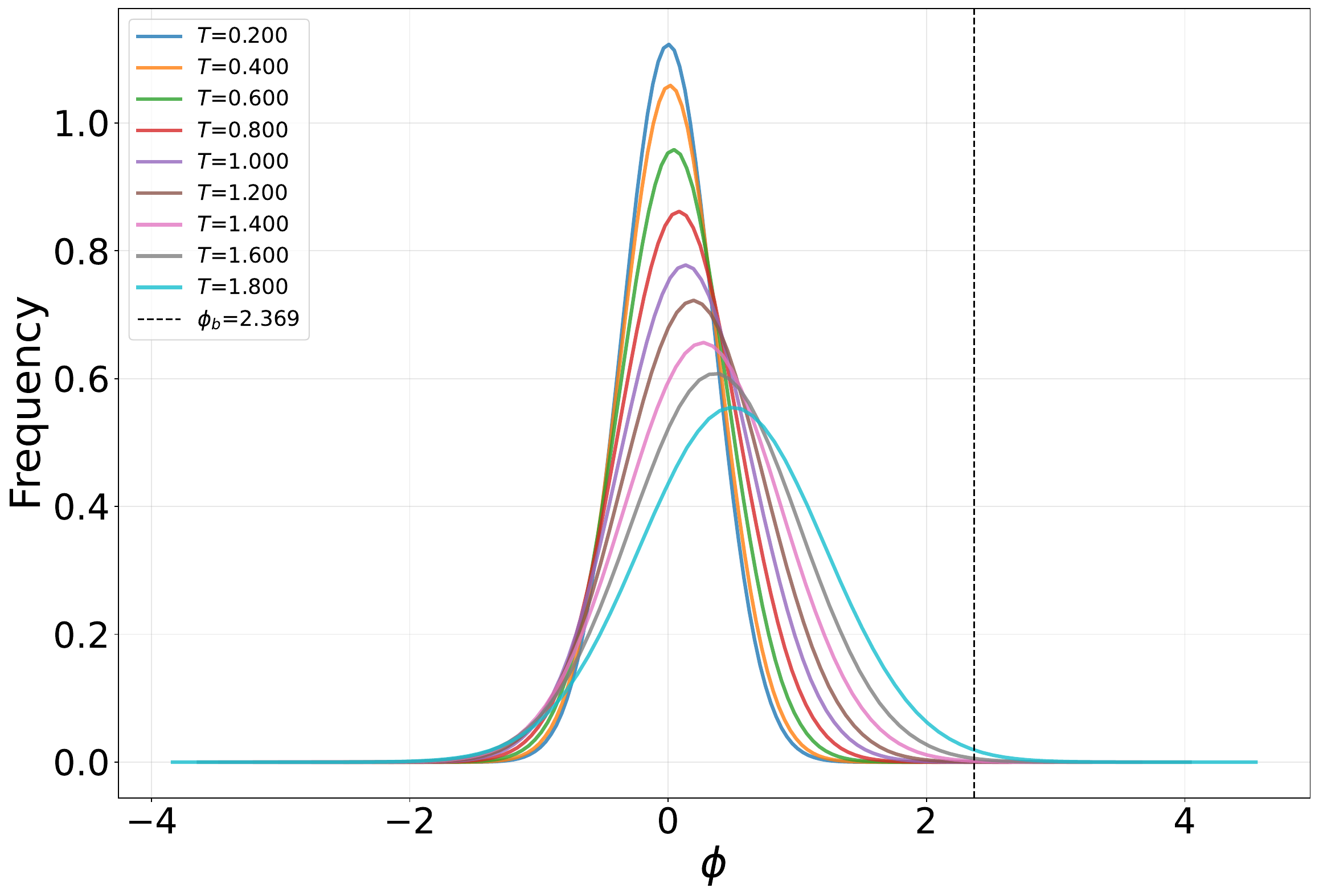}
    \includegraphics[width=0.3\linewidth]{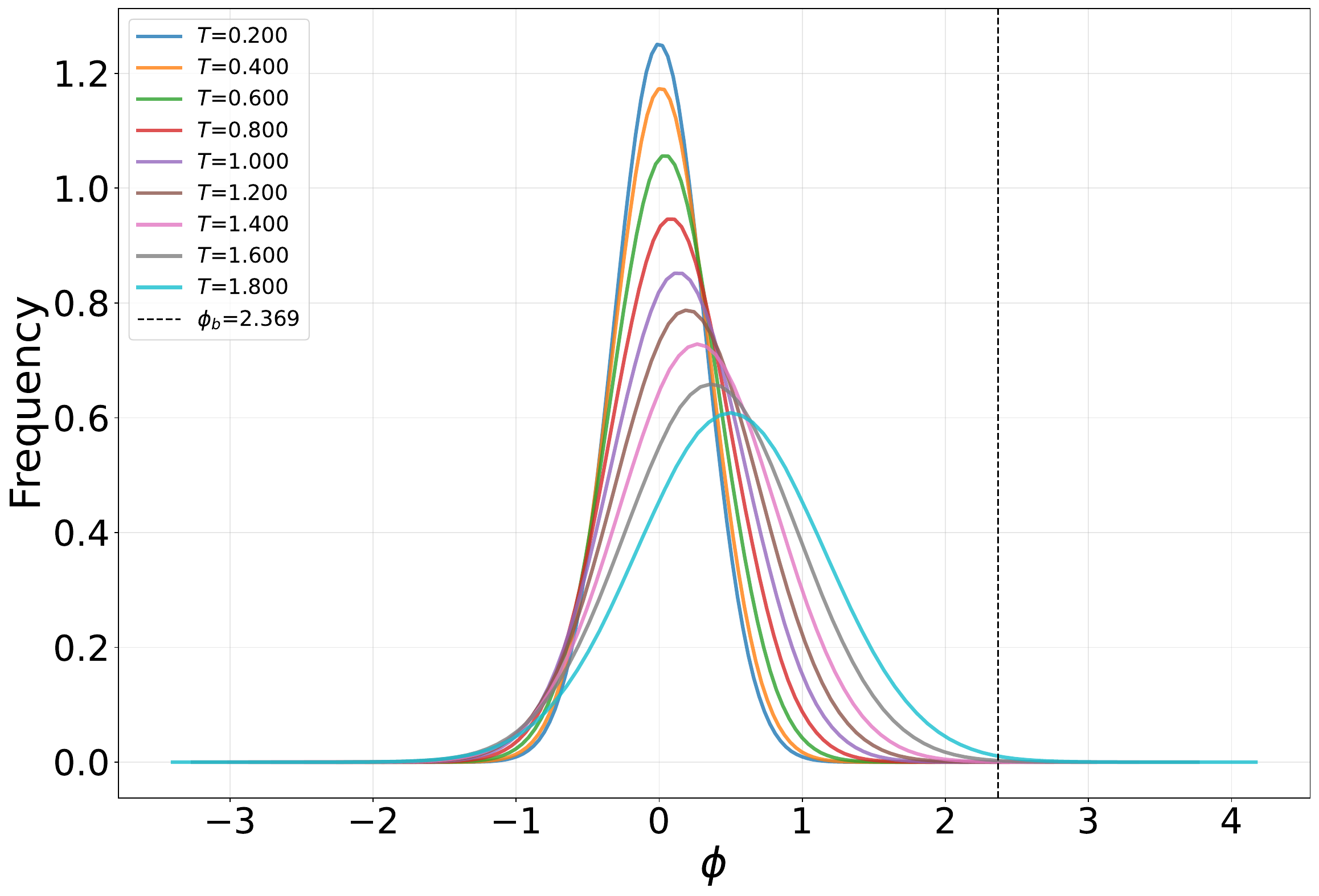}
    \caption{Initial coarse-grained field-value distributions for different temperatures. From left to right, the coarse-graining coefficient is $c_1=2,3,4$. The dashed vertical line denotes the barrier-top field value $\phi_b$.}
    \label{fig:init_phi}
\end{figure}

The dependence of the initial coarse-grained distribution on $c_1$ is therefore natural and unavoidable: the plotted quantity is itself resolution dependent. What matters physically is not whether the histogram width changes under coarse graining, but whether the decay observables extracted from the corresponding coarse-grained field remain stable over a reasonable range of $c_1$. In practice, although the initial coarse-grained field distribution changes visibly as $c_1$ varies, the final extracted decay rates show only weak dependence on $c_1$ within the range considered in this work, as shown in the parameter dependence analysis below. This is the behavior expected from a sensible coarse-graining prescription: short-distance diagnostics depend on resolution, while long-time decay observables approach a stable regime.

For this reason, we use $c_1=3$ as the representative benchmark in the main text. This value provides a reasonable compromise between suppressing ultraviolet pointwise noise and retaining the physically relevant domain structure. Smaller values of $c_1$ leave more short wavelength fluctuations in the analysis field, whereas larger values over-smooth the field and reduce spatial resolution without significantly improving the stability of the extracted decay rate.

\begin{figure}[H]
    \centering
    \includegraphics[width=0.46\linewidth]{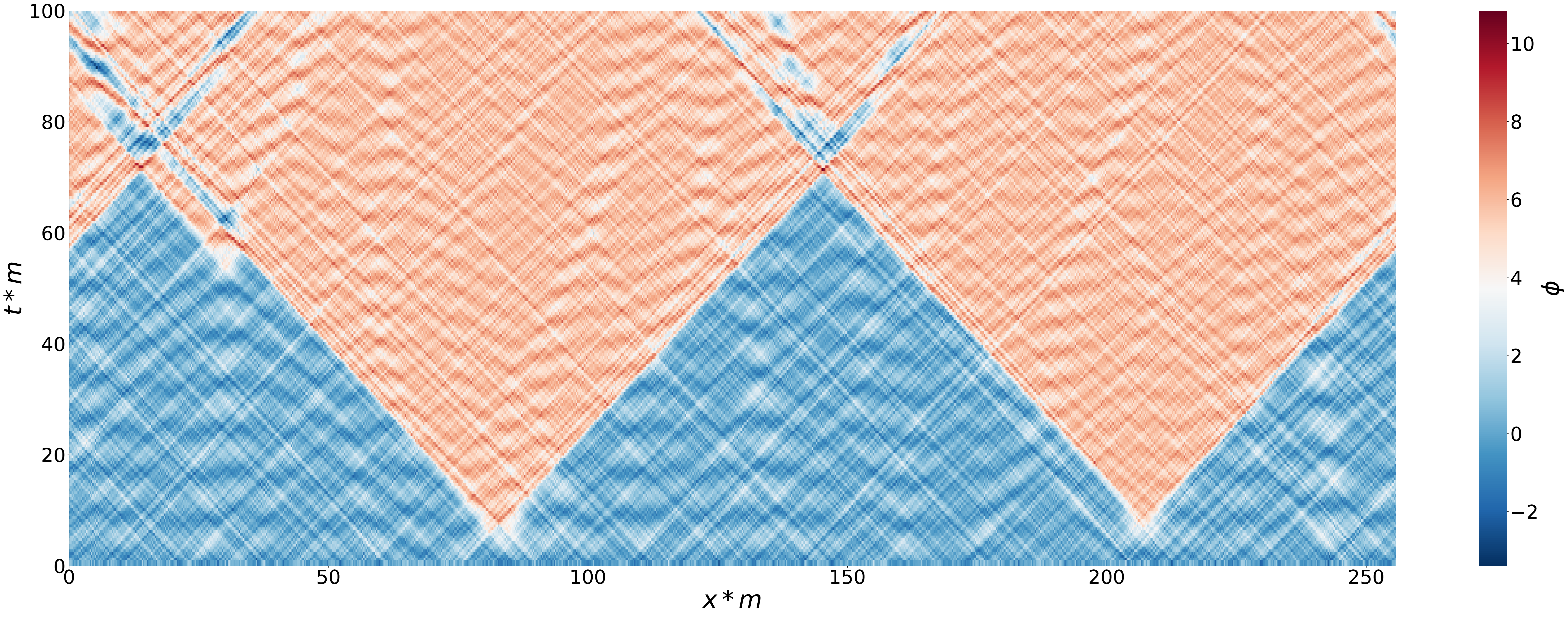}
    \includegraphics[width=0.46\linewidth]{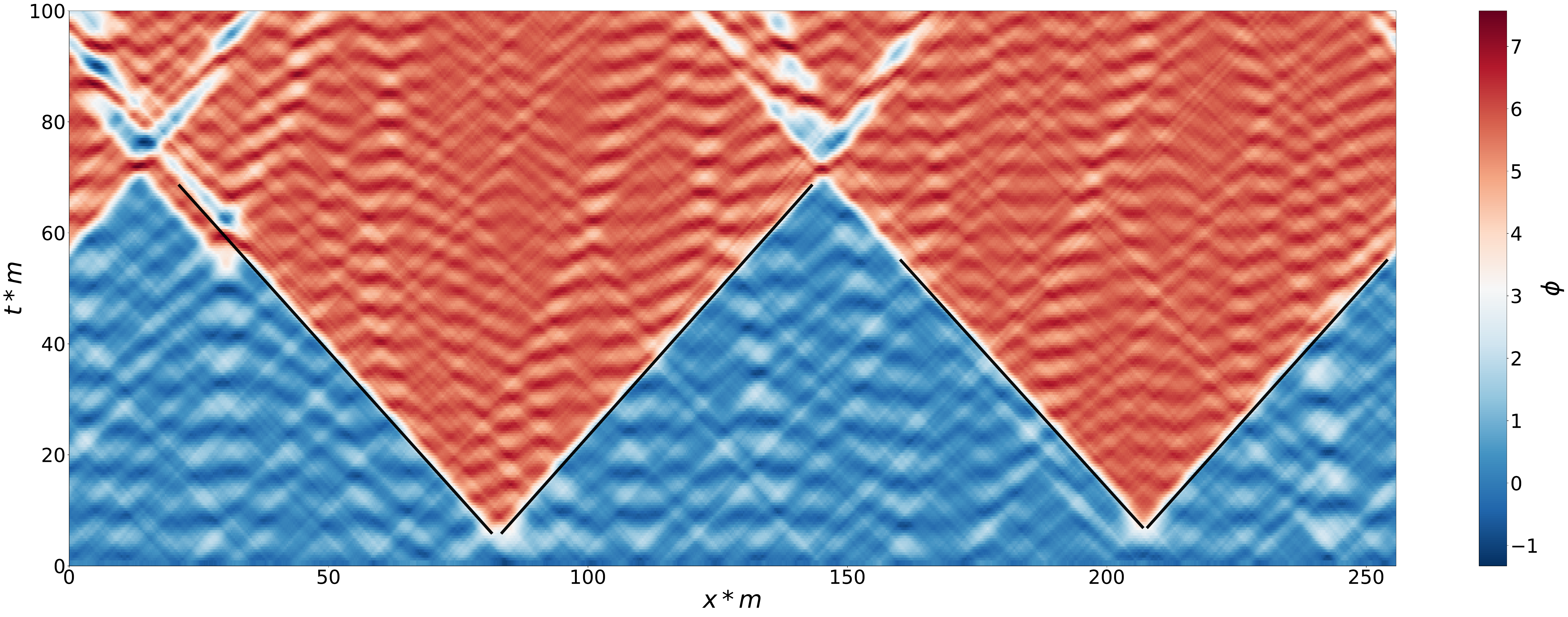}
    \includegraphics[width=0.46\linewidth]{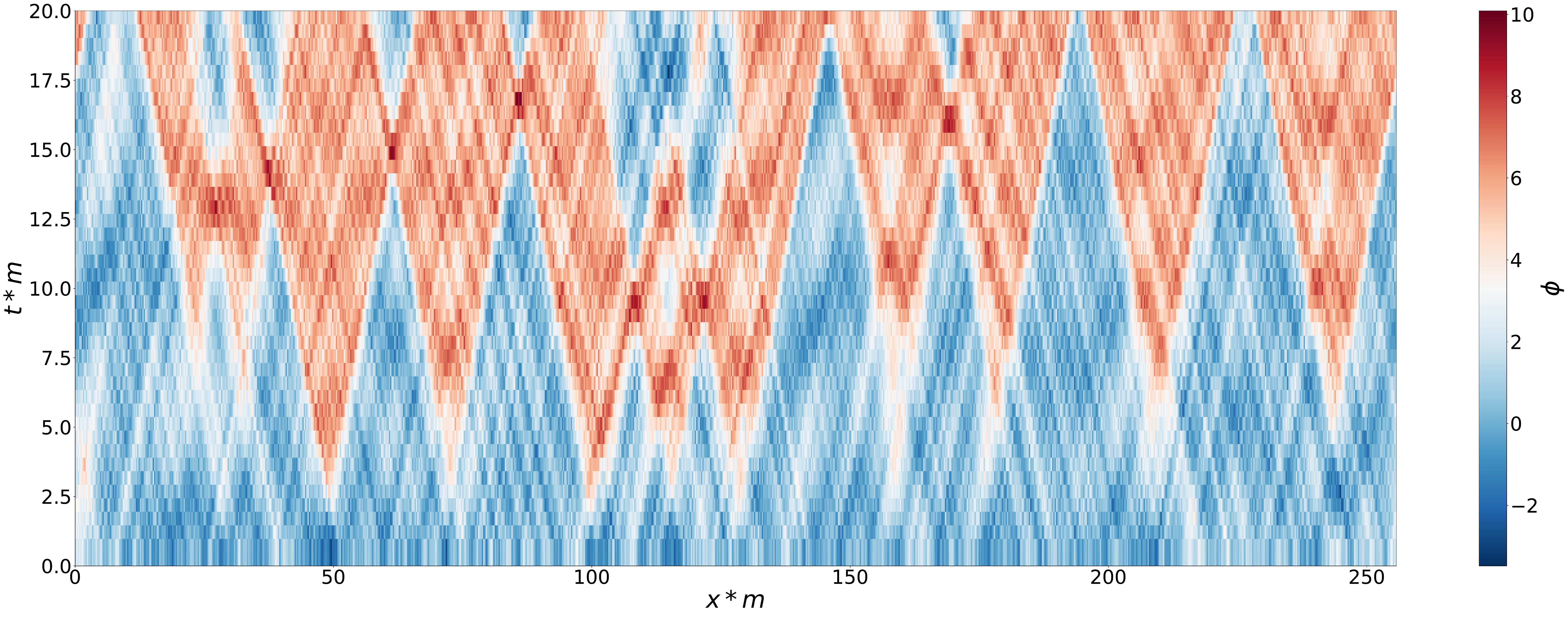}
    \includegraphics[width=0.46\linewidth]{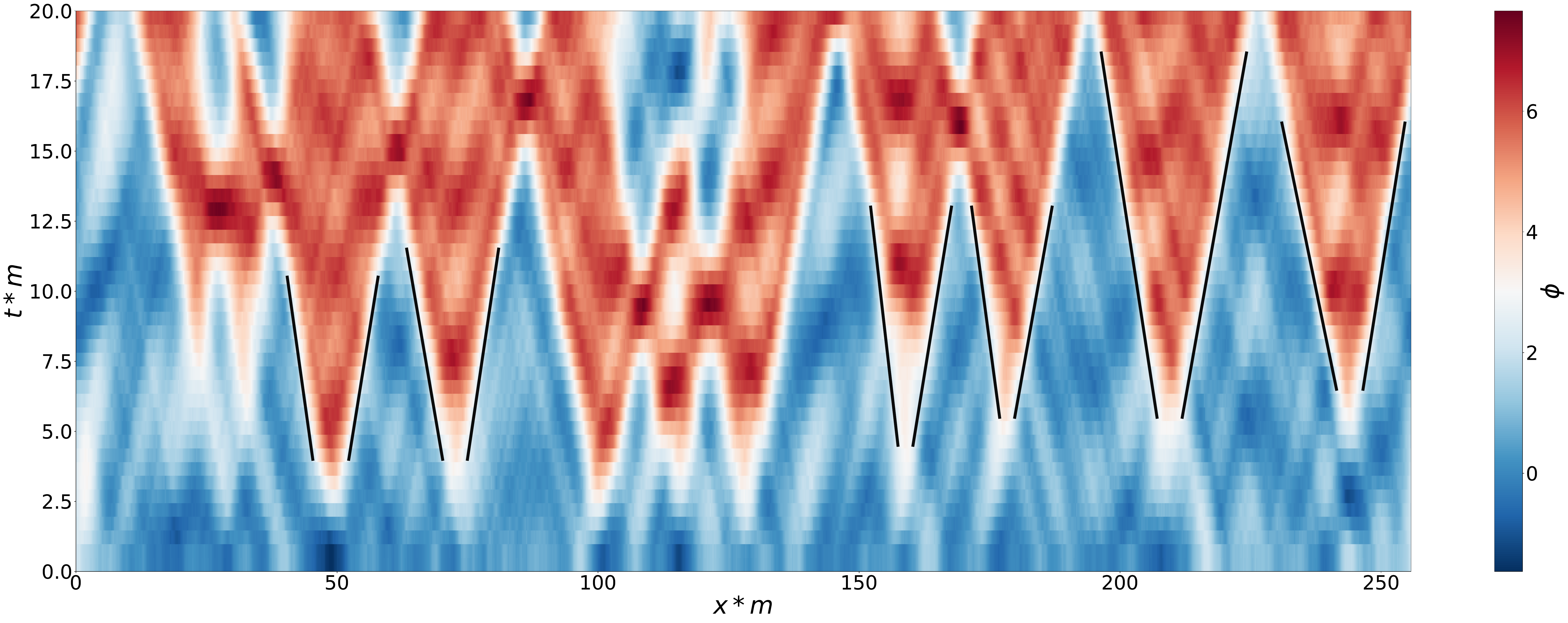}
    \caption{Comparison of spacetime evolution diagrams for randomly selected sample fields. The upper panels correspond to $T=0.4$, while the lower panels correspond to $T=1.6$. In each row, the left panel shows the original field and the right panel shows the corresponding coarse-grained field.}
    \label{fig:field_evol_coarse_compare}
\end{figure}

Figure~\ref{fig:field_evol_coarse_compare} illustrates the effect of the box-smoothing procedure on representative real time field configurations. The coarse-grained profiles remove rapid small scale oscillations while preserving the larger connected structures associated with true vacuum domain formation. This comparison provides a useful visual check that coarse graining acts as an observational filter: it suppresses short-wavelength noise but does not alter the underlying microscopic evolution. The black lines in the coarse-grained panels indicate representative domain wall trajectories used to estimate the kink-antikink propagation speed in one spatial dimension.

The upper panels of Fig.~\ref{fig:field_evol_coarse_compare} show a representative low-temperature event at $T=0.4$. In this sample, two bubble-like domains appear around $x\simeq 80$ and $x\simeq 210$ and subsequently expand. In one spatial dimension, these expanding domains correspond to kink-antikink domain wall pairs separating the false-vacuum and true-vacuum regions. As the two domains expand toward each other, the false-vacuum region between them is squeezed. Around $t\simeq 70$, the antikink from the left domain and the kink from the right domain collide near the middle of the lattice. After the collision, the domain walls move back toward their original sides, indicating a partial reflection or return dynamics rather than an irreversible conversion of the intervening region. This behavior illustrates that, in the low-temperature regime, a region that has temporarily crossed to the true-vacuum side need not remain permanently converted. Such partial-return or reflection dynamics is precisely the type of effect that can make a purely fraction-based observable difficult to interpret as a sample-level decay probability. The fraction observable counts transient spatial conversion, whereas the connected-cluster survival criterion requires a sufficiently large domain to persist for a holding time and is therefore less sensitive to short-lived threshold crossings.

The lower panels of Fig.~\ref{fig:field_evol_coarse_compare} show a representative high-temperature event at $T=1.6$. In this case, thermal fluctuations generate several local transformed regions, and the coarse-grained domains form and grow on a shorter time scale. The dynamics is less dominated by coherent kink-antikink reflection and is closer to an irreversible nucleation-and-growth picture. This provides a useful qualitative reason why the fraction-based Avrami analysis becomes more meaningful at higher temperature, while the connected-cluster criterion remains a more direct sample-level diagnostic of persistent supercritical-domain formation.

\begin{figure}[!htp]
    \centering
    \includegraphics[width=0.46\linewidth]{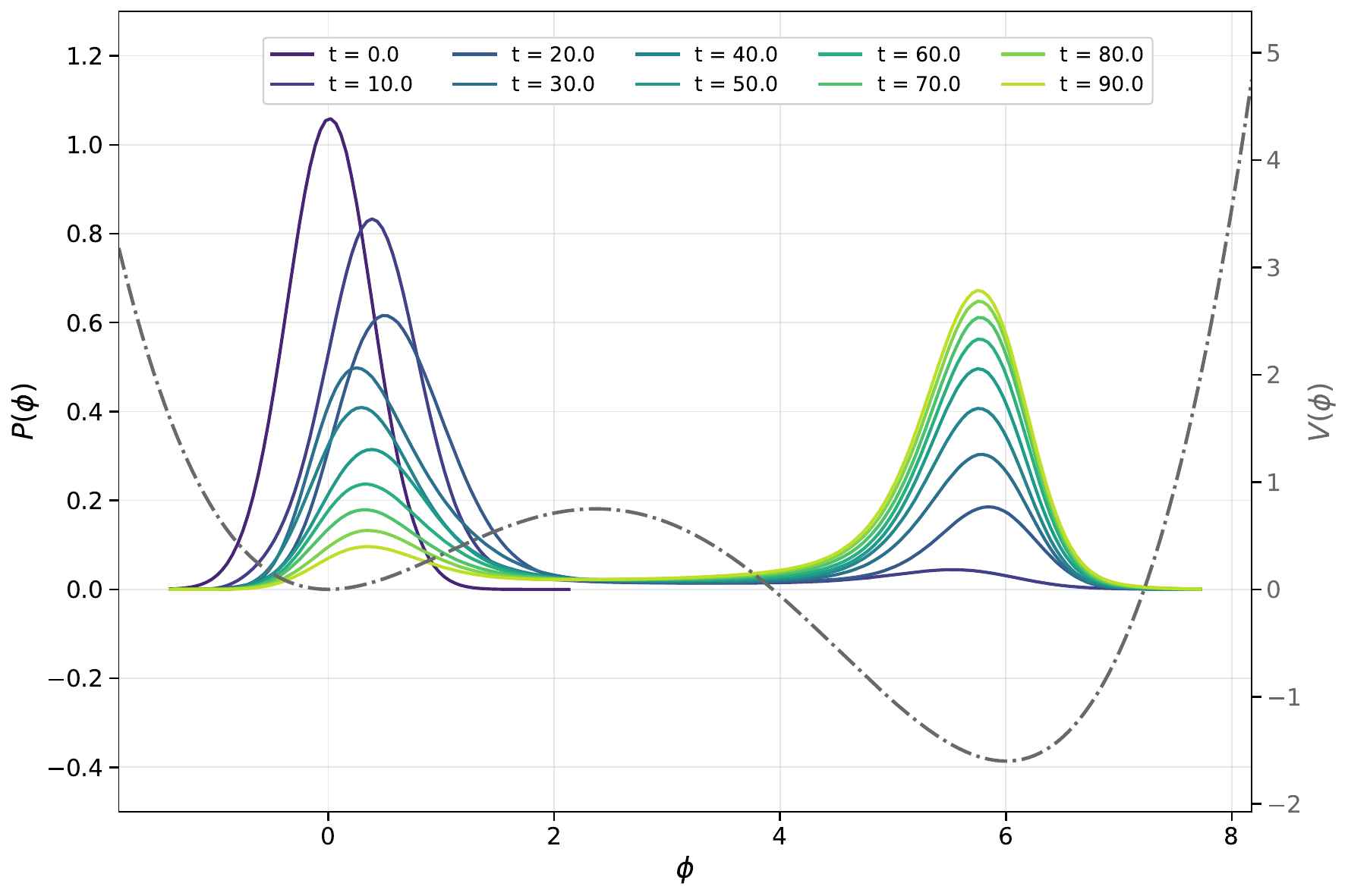}
    \includegraphics[width=0.46\linewidth]{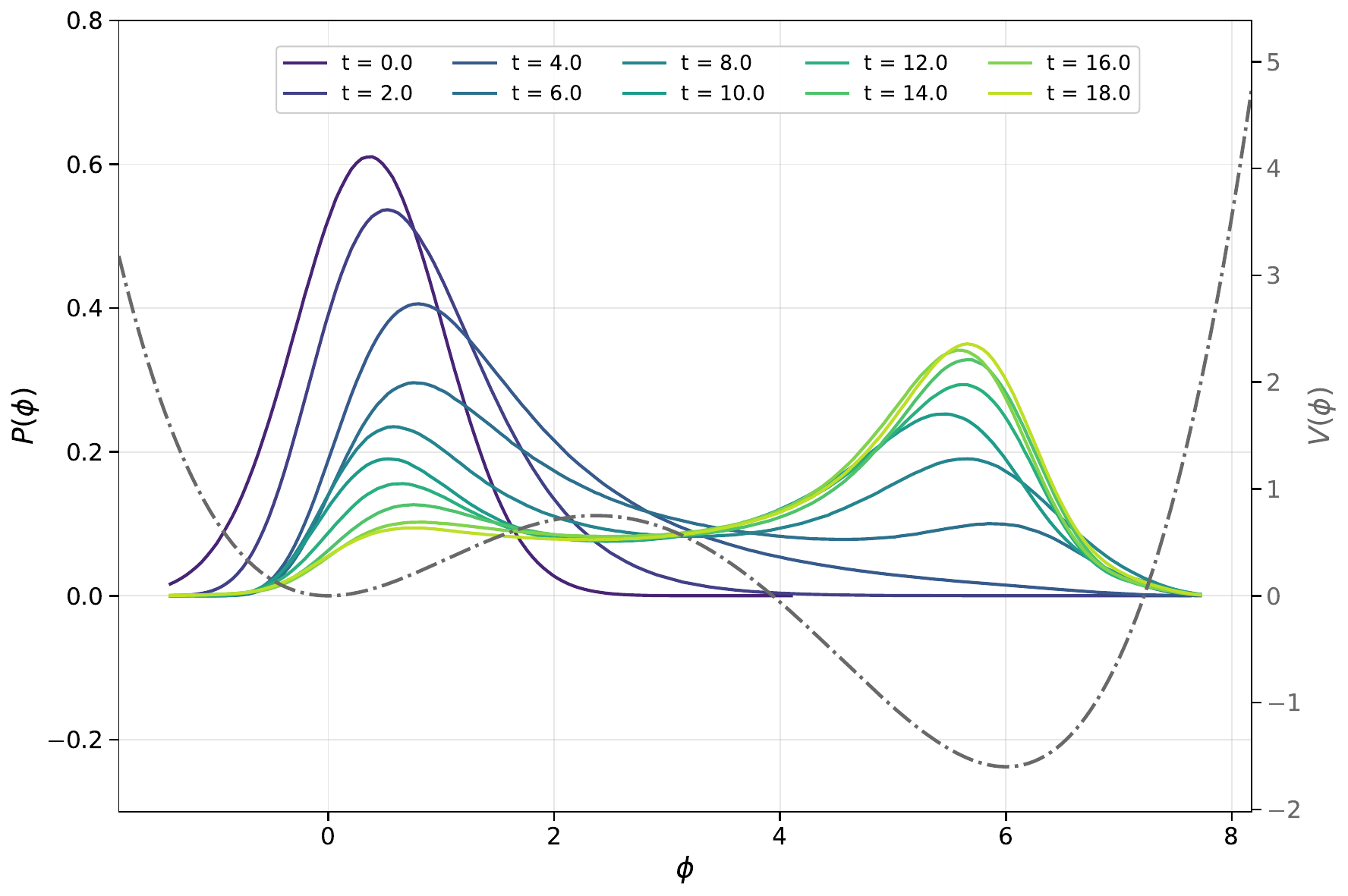}
    \caption{Evolution of the projected coarse-grained field value distribution($c_1=3$). The left panel corresponds to $T = 0.4$, and the right panel corresponds to $T = 1.6$. The gray dashed curve shows the bare potential as a guide to the field value landscape.}
    \label{fig:probability_dis_evolve}
\end{figure}
% By dividing the field values into 200 sub-intervals, we can calculate the probability of each sub-interval using the similar way as Eq. (15, 16) in the main text. Eventually, we obtain the quasi-probability distribution evolution graph shown in Fig.~\ref{fig:probability_dis_evolve}. Admittedly, we cannot interpret this as a true quantum state probability distribution. However, as a realization of functional integration, Eq. (1), the quasi-probability distribution obtained through this method can still be regarded as a measurable quantity for the decay process. 
% % From the diagrams here, we can also observe that the evolution behavior of the quasi-probability density varies under different temperatures ?.
We also examine the time evolution of the projected field value distribution, shown in Fig.~\ref{fig:probability_dis_evolve}. To construct this diagnostic, we divide the field value range into 200 bins and compute the fraction of coarse-grained lattice points, averaged over the ensemble, that fall into each bin at a given time. This is analogous to the local-indicator averaging used in Eqs.~(15) and (16) of the main text, but here the full field value histogram is retained rather than only the false vacuum fraction. This quantity should not be interpreted as the full quantum probability distribution or as the full Wigner functional. It is instead a one dimensional projection of the sampled real-time ensemble onto the coarse-grained field value.

The projected distribution is nevertheless useful for diagnosing the decay process. At $T=0.4$, the distribution develops a second peak on the true vacuum side while retaining a pronounced false-vacuum peak, reflecting the coexistence of false vacuum regions and transformed domains as well as the partial return dynamics visible in the spacetime plots. At $T=1.6$, the distribution shifts toward the true-vacuum side on a much shorter time scale, indicating faster spatial conversion driven by stronger thermal fluctuations. Thus Fig.~\ref{fig:probability_dis_evolve} provides a compact visualization of how the ensemble gradually leaves the false-vacuum basin and how this process differs between the low- and high-temperature regimes.

The same coarse-grained sample profiles can also be used to estimate the bubble-wall velocity entering the Avrami-type fit in the main text. In practice, for each saved coarse-grained field profile we first apply a short smoothing window and then identify connected segments satisfying $\phi(x,t)>\phi_m$, where $\phi_m$ is the field value at the barrier top. A candidate bubble is retained only if its width and peak height exceed minimal quality cuts, so that short-lived noise spikes are excluded. The left and right boundaries of each accepted segment are determined by interpolation at the threshold crossing, and the corresponding bubble tracks are then followed in time by matching neighboring segments with nearby centers under periodic boundary conditions. For each sufficiently long and steadily growing track, we fit the left and right boundary positions separately as linear functions of time. The bubble velocity is defined as the average of the absolute values of the fitted left- and right-wall velocities,
\begin{equation}
v_{\rm bubble}=\frac{|v_{\rm left}|+|v_{\rm right}|}{2}.
\end{equation}
Tracks with poor linearity or strong left-right asymmetry are discarded.

The resulting temperature dependence of the extracted bubble velocity is shown in Fig.~\ref{fig:bubble_velocity_vs_T}. In the temperature range considered here, the median bubble velocity remains close to unity and exhibits only mild temperature dependence, while the error bars indicate the spread among different accepted bubble tracks. This justifies the use of a nearly constant relativistic wall speed as an effective input in the Avrami-type analysis, while still allowing for a quantitative uncertainty estimate from the sample-to-sample variation.

\begin{figure}[!htp]
    \centering
    \includegraphics[width=0.8\linewidth]{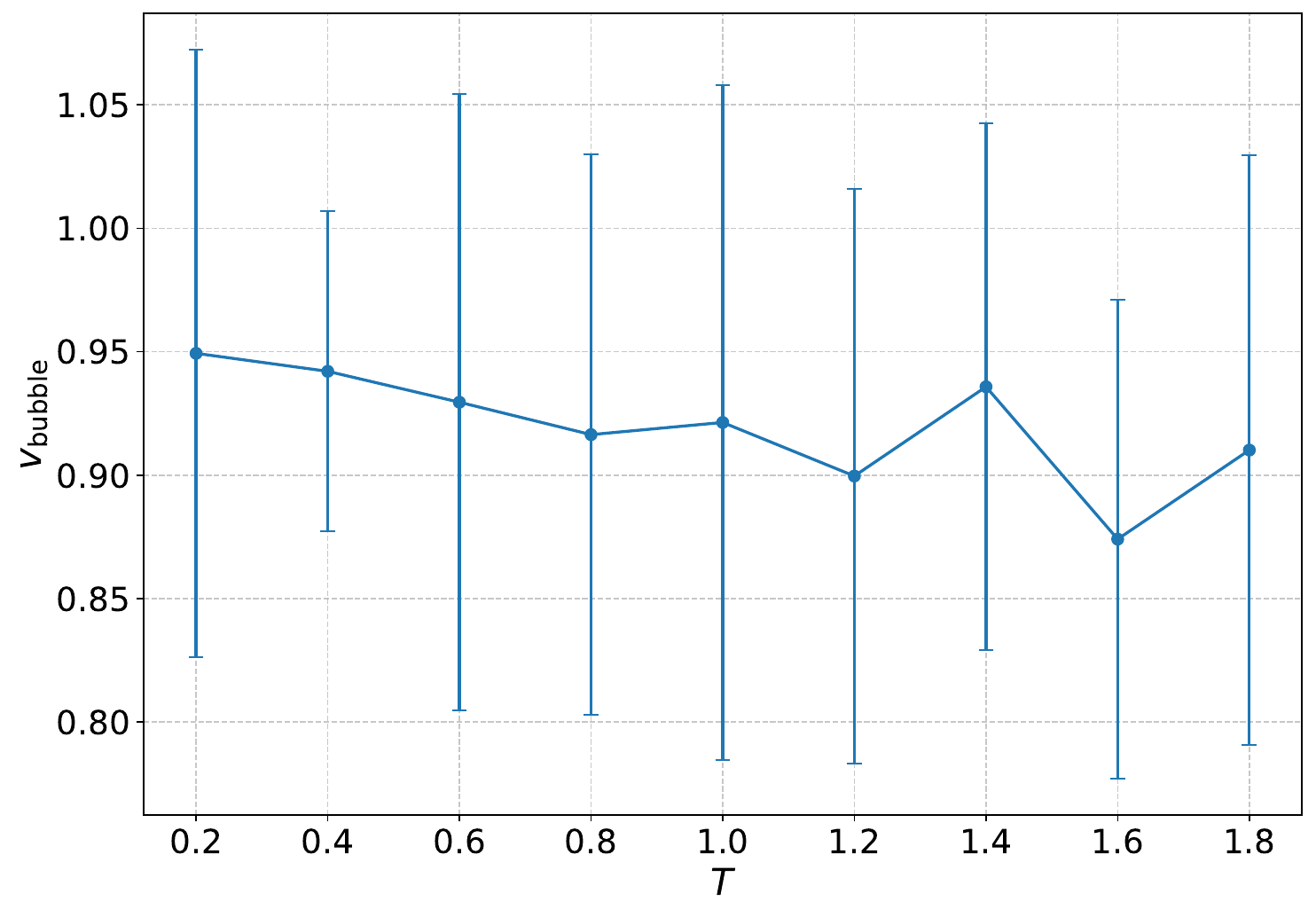}
    \caption{Bubble velocity extracted from coarse-grained sample field evolutions at different temperatures. The central value at each temperature is taken to be the median over accepted bubble tracks, and the error bar shows the corresponding standard deviation.}
    \label{fig:bubble_velocity_vs_T}
\end{figure}

\section{Hartree background effective potential and thermal nucleation benchmark}
\label{app:hartree_veff}

To compare our real time results with the standard finite temperature nucleation picture, we construct a temperature dependent Hartree background effective potential using the same renormalization convention as in the initial state calculation. The purpose of this construction is not to redefine the real time dynamics, but to provide a self consistent thermal benchmark potential from which a semiclassical nucleation rate can be estimated.

In the Hartree-Gaussian construction of the initial ensemble discussed in Sec.~\ref{app:hartree_init}, the self-consistent stationary background $\varphi_\star$ is determined together with the fluctuation variance $G$ and the effective mass $M_H$. Here, in order to reconstruct the full background free energy landscape, we instead treat the uniform background field $\varphi$ as an external scanning parameter. Specifically, we discretize the interval
\begin{equation}
\varphi\in[\varphi_{\min},\varphi_{\max}]
\end{equation}
into a dense grid and, at each fixed $\varphi$, solve the Hartree gap equation for the corresponding fluctuation variance $G_{\rm eff}(\varphi,T)$ and effective mass $M_H^2(\varphi,T)$:
\begin{equation}
M_H^2(\varphi,T)=m^2+2g\varphi+3\lambda\varphi^2+3\lambda G_{\rm eff}(\varphi,T),
\label{eq:MH_phiT}
\end{equation}
\begin{equation}
G_{\rm eff}(\varphi,T)
=
\frac{1}{L}\sum_k
\left[
\frac{n_B(\omega_k)+1/2}{\omega_k}
\right]
\qquad
\text{or its subtracted form},
\label{eq:G_phiT}
\end{equation}
with
\begin{equation}
\omega_k^2=\hat{k}^2+M_H^2(\varphi,T).
\end{equation}
In the actual implementation, we use the same subtracted scheme as in the initial-condition code, so that the background effective potential and the Hartree initial ensemble share the same renormalization convention.

At fixed $\varphi$, the derivative of the Hartree background effective potential is then reconstructed from the tadpole coefficient:
\begin{equation}
\frac{\partial V_{\rm eff}(\varphi;T)}{\partial \varphi}
=
m^2\varphi+g\varphi^2+\lambda\varphi^3
+g\,G_{\rm eff}(\varphi,T)
+3\lambda\varphi\,G_{\rm eff}(\varphi,T).
\label{eq:dveff_dphi_hartree}
\end{equation}
The effective potential itself is obtained by numerical integration,
\begin{equation}
V_{\rm eff}(\varphi;T)
=
\int^{\varphi} d\varphi'\,
\frac{\partial V_{\rm eff}(\varphi';T)}{\partial \varphi'}
+ {\rm const},
\label{eq:veff_reconstruct}
\end{equation}
where the additive constant is fixed by shifting the false-vacuum-side local minimum to zero. In our numerical scan, we take $\varphi_{\min}=-1,~\varphi_{\max}=8$, and discretize this interval with 801 grid points.

The resulting Hartree background effective potentials are shown in Fig.~\ref{fig:hartree_veff}. As the temperature increases, the effective barrier is gradually lowered and the metastable minimum becomes less pronounced. This trend is consistent with the intuitive expectation that thermal fluctuations reduce the free-energy cost of forming a supercritical configuration. At sufficiently high temperature, the effective barrier becomes very shallow, indicating that the system approaches the near-spinodal regime.

\begin{figure}[!htp]
    \centering
    \includegraphics[width=0.8\linewidth]{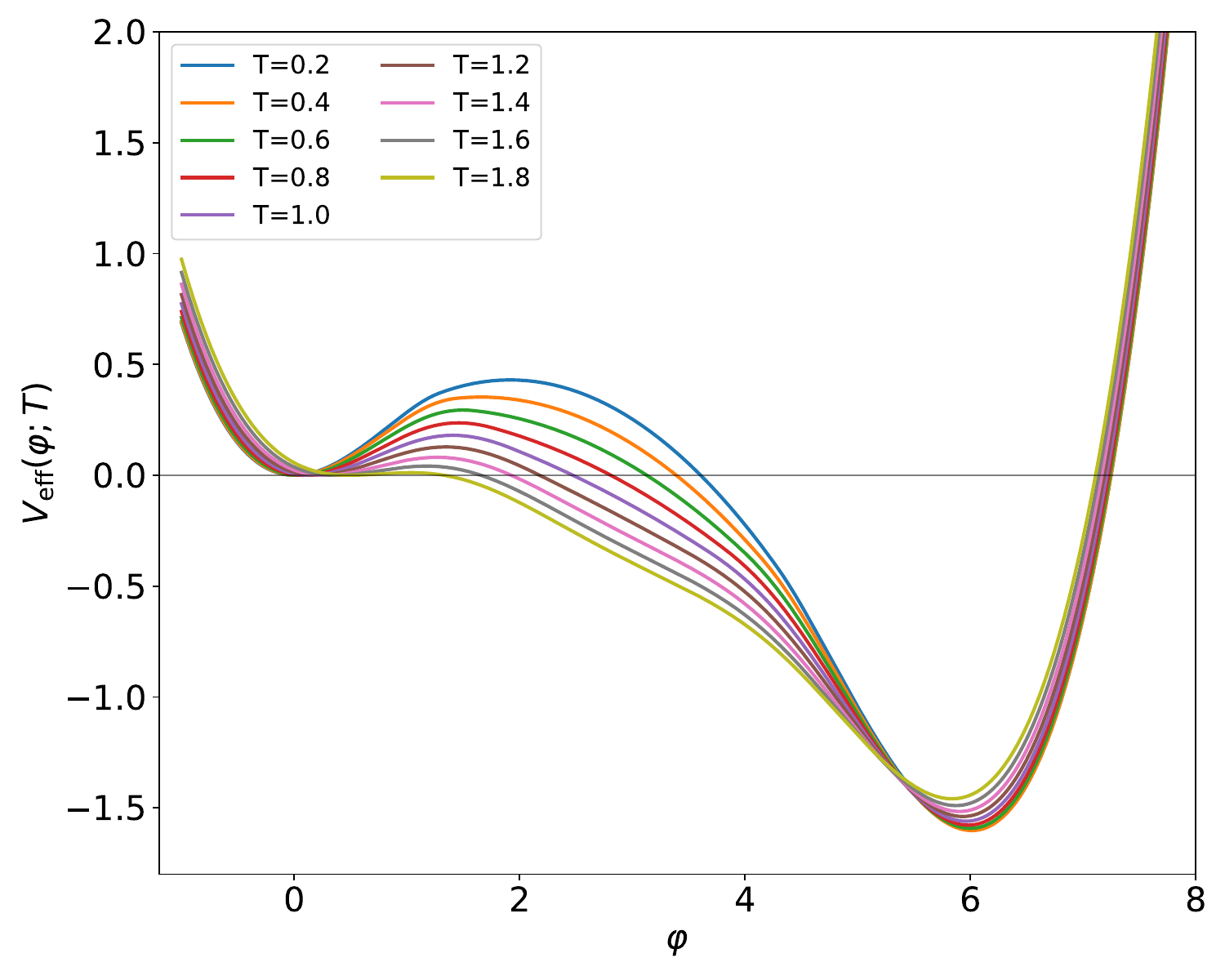}
    \caption{The Hartree background effective potential reconstructed from Eq.~(\ref{eq:dveff_dphi_hartree}) at different temperatures, the false vacuum side local minimum is shifted to zero for each temperature. }
    \label{fig:hartree_veff}
\end{figure}

To obtain a theoretical benchmark for thermally activated decay, we feed the reconstructed $V_{\rm eff}(\varphi;T)$ into FindBounce and BubbleDet and solve for the corresponding critical bubble / bounce configuration. This gives the semiclassical thermal benchmark rate associated with the Hartree effective potential.
% This gives the semiclassical thermal nucleation rate per unit volume of the standard form:
% \begin{equation}
% \Gamma_{\rm th}
% \sim
% A(T)\exp\!\left[-\frac{\Delta F_{\rm cb}(T)}{T}\right],
% \label{eq:thermal_benchmark_rate}
% \end{equation}
% where $\Delta F_{\rm cb}(T)$ is the free-energy cost of the critical bubble obtained from the Hartree effective potential, and $A(T)$ denotes the corresponding prefactor returned by the solver.

At high temperature, however, the interpretation of this benchmark requires some care. Once the barrier becomes too shallow, the separation between the false vacuum, the barrier top, and the true-vacuum side is no longer sharp, and both the polynomial reconstruction of $V_{\rm eff}$ and the semiclassical bounce calculation become increasingly sensitive to numerical details. In our results, this is the origin of the apparent turnover of the theoretical curve at $T\gtrsim 1.4$: it should be regarded as a signal that the effective barrier is approaching disappearance and that the semiclassical benchmark is losing reliability in this regime, rather than as a robust physical suppression of the decay rate.

\section{Fitting prescription and parameter dependence of real-time observables}
\label{app:fitting_parameter_dependence}

In this work, the false-vacuum probability is evaluated using several implementations of the indicator functional $\Theta_{FV}$, including the global-average survival criterion, the connected-cluster survival criterion, and the fraction-based observable. These observables probe different aspects of the real time decay dynamics and therefore require different fitting prescriptions. In this section, we describe the fitting window selection used to extract the effective decay rates and then examine the dependence of the results on the coarse-graining and indicator parameters.

The extraction of an effective decay rate requires an intermediate time fitting window. Very early times are affected by the transient relaxation of the sampled initial ensemble, short-lived threshold crossings, and, for the connected-cluster criterion, the finite holding time requirement. Very late times are affected by finite-sample noise, since only a small number of realizations remain undecayed. We therefore select the fitting window by examining the approximate linearity of the transformed variables appropriate to each observable.

For the survival type observables, we fit $\ln P_{\rm surv}(t)=-\Gamma_{\rm surv}Lt+C$ only in the region where $\ln P_{\rm surv}$ is approximately linear in $t$. In practice, the connected-cluster survival curve can show a two stage structure at low and intermediate temperatures. The initial rapid decrease is interpreted as the removal of configurations containing near critical or short lived supercritical clusters already present in the fluctuating initial ensemble, together with the gating effect introduced by the holding time requirement. The later stage, when it exists over a sufficiently long interval and remains above the finite-sample tail, is used to extract the Poisson-like sample-level decay rate.

For the fraction-based observable, the appropriate transformed variable is $\ln P_{\rm frac}$ plotted against $t^2$: $\ln P_{\rm frac}(t)=-\Gamma_{\rm frac}v t^2+C$. The fitting window is chosen where this Avrami transformed relation is approximately linear. We avoid very late times, where bubble overlap, post collision oscillations, and partial returns toward the false vacuum side can distort the transformed spatial fraction. To estimate the fitting uncertainty, we scan smaller subwindows inside the admissible fitting range and report the spread of the resulting slopes. The same window selection logic is applied when varying the coarse-graining and indicator parameters.

The real time observables depend on four auxiliary parameters introduced in the main text. The role of these parameters is different. The parameter $c_1$ controls the observational resolution of the coarse-grained field, $c_2$ controls how far beyond the barrier a local field value must move before being counted as transformed, $c_3$ sets the minimum size of a connected supercritical interval, and $c_4$ determines how long such an interval must persist before the sample is declared to have decayed.

We first examine the dependence on the coarse-graining coefficient $c_1$. Although the initial coarse-grained field distribution depends visibly on the smoothing scale, the real time observables relevant for the decay analysis are much less sensitive to $c_1$ within the range considered here. This is shown in Figs.~\ref{fig:app_c1_psurv} and \ref{fig:app_c1_pfrac}, where the connected cluster survival probability and the fraction based observable are compared for several coarse-graining lengths. The weak dependence of these observables on $c_1$ supports the interpretation of the coarse-graining scale as a resolution parameter: it suppresses short wavelength pointwise noise without changing the large scale decay dynamics over a reasonable range.

\begin{figure}[H]
    \centering
    \includegraphics[width=0.22\linewidth]{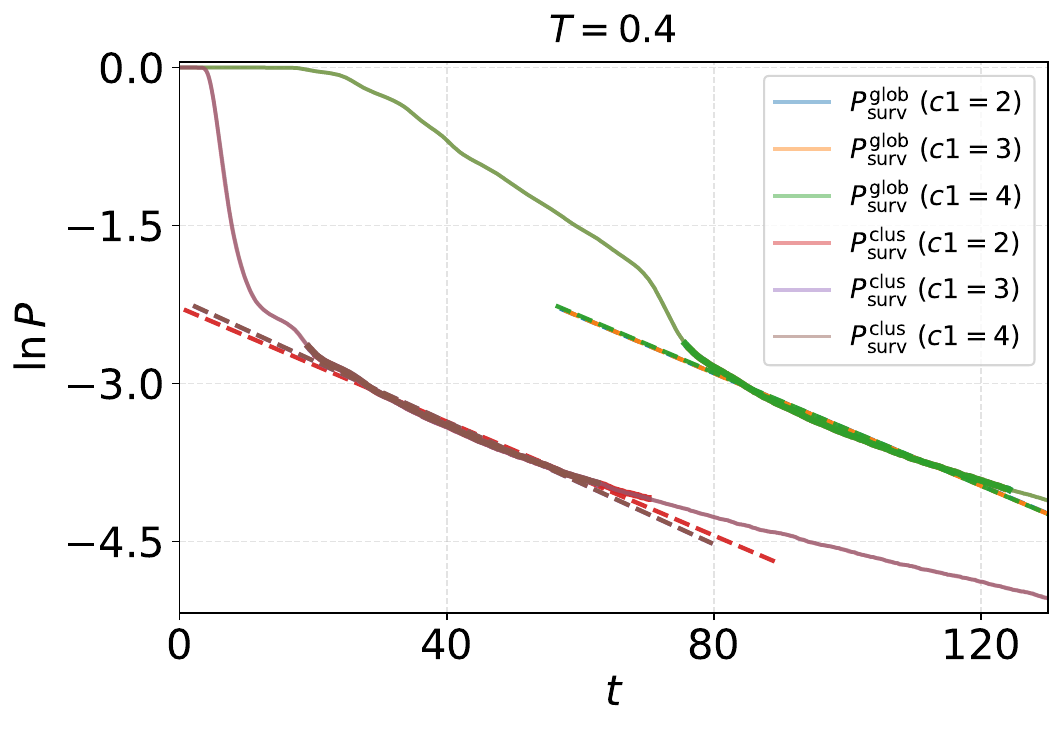}
    \includegraphics[width=0.22\linewidth]{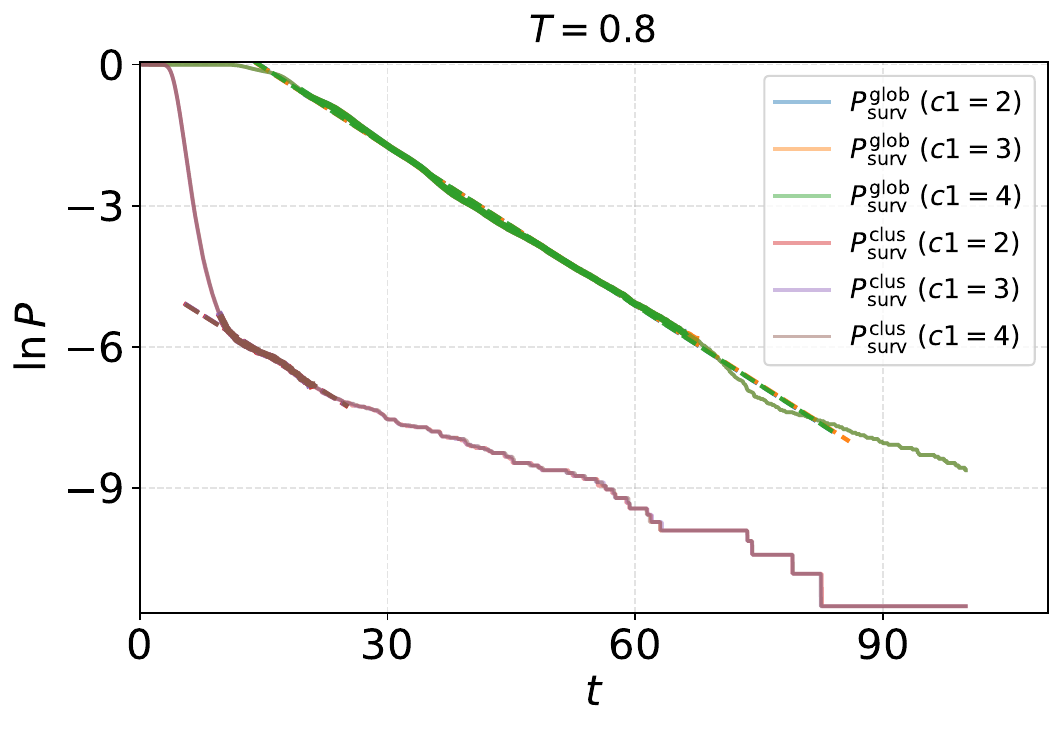}
    \includegraphics[width=0.22\linewidth]{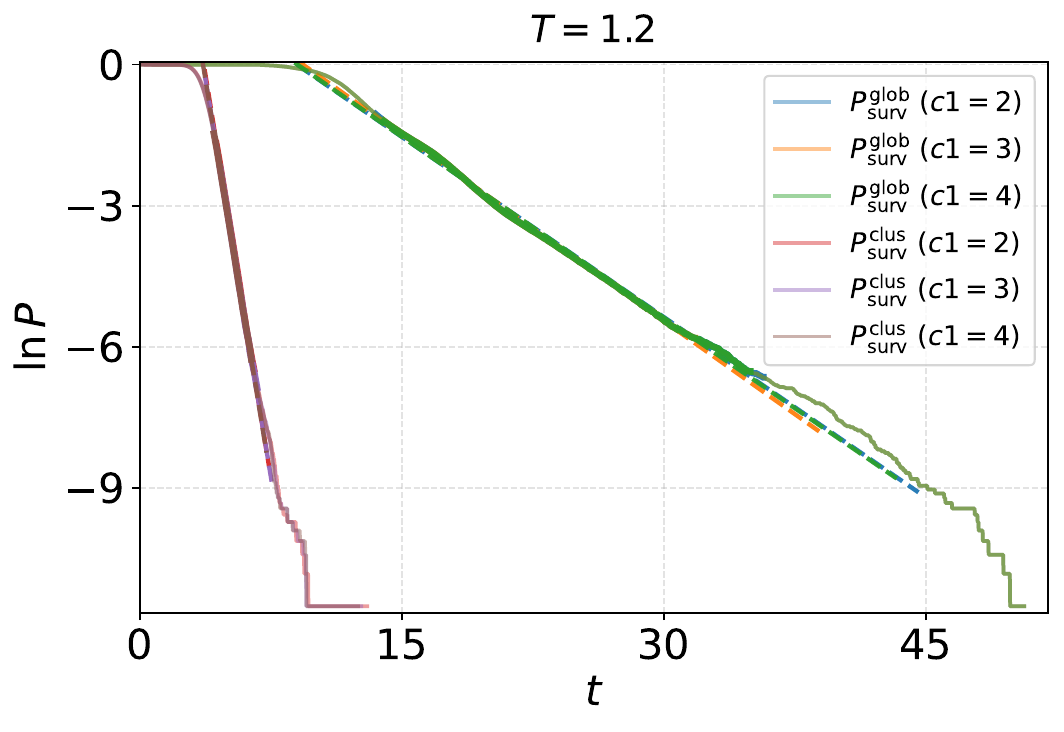}
    \includegraphics[width=0.22\linewidth]{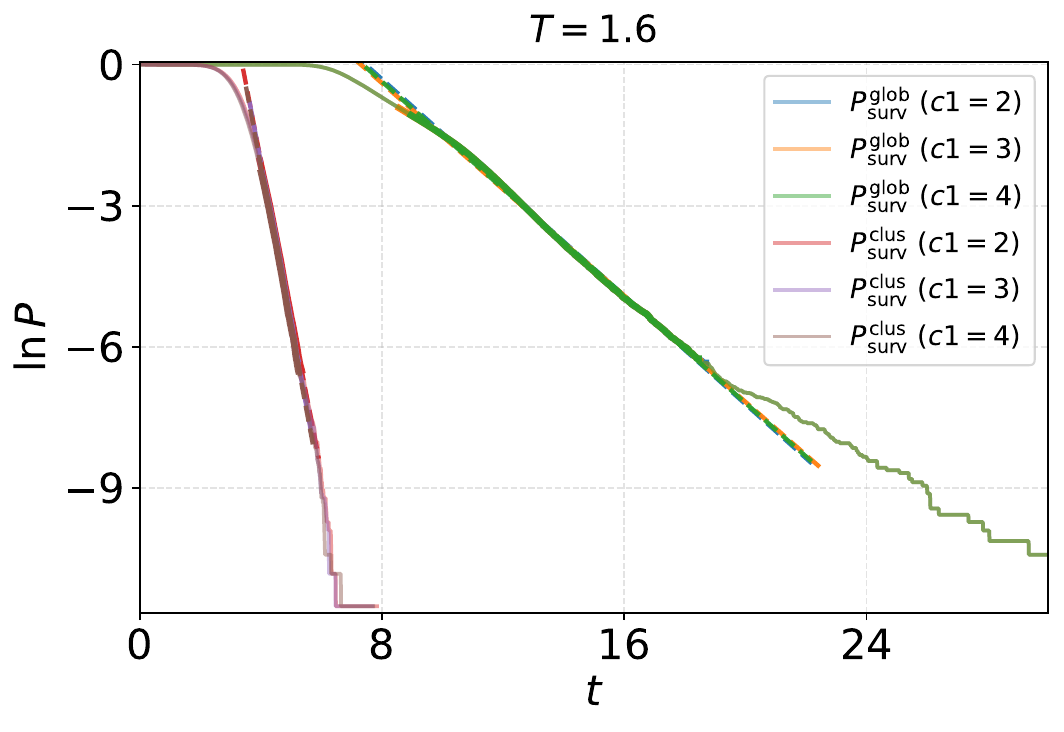}
    \caption{Time evolution of the survival probability for different coarse-graining lengths. Here $c_2=0.8$, $c_3=5$, and $c_4=3$.}
    \label{fig:app_c1_psurv}
\end{figure}

\begin{figure}[H]
    \centering
    \includegraphics[width=0.22\linewidth]{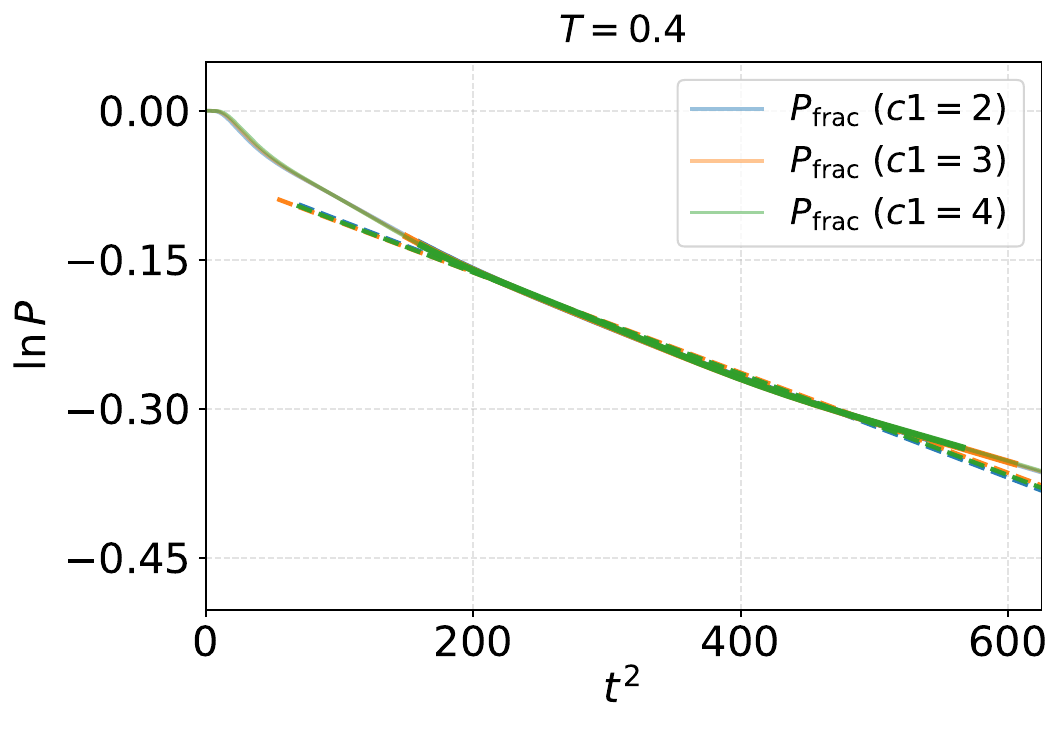}
    \includegraphics[width=0.22\linewidth]{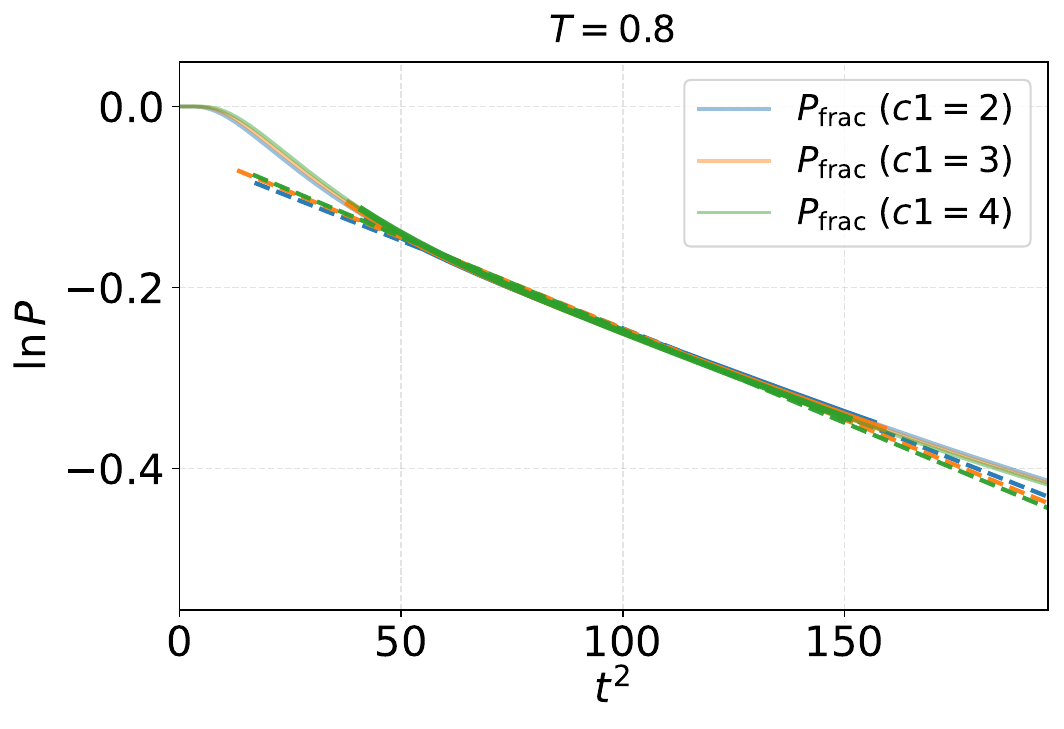}
    \includegraphics[width=0.22\linewidth]{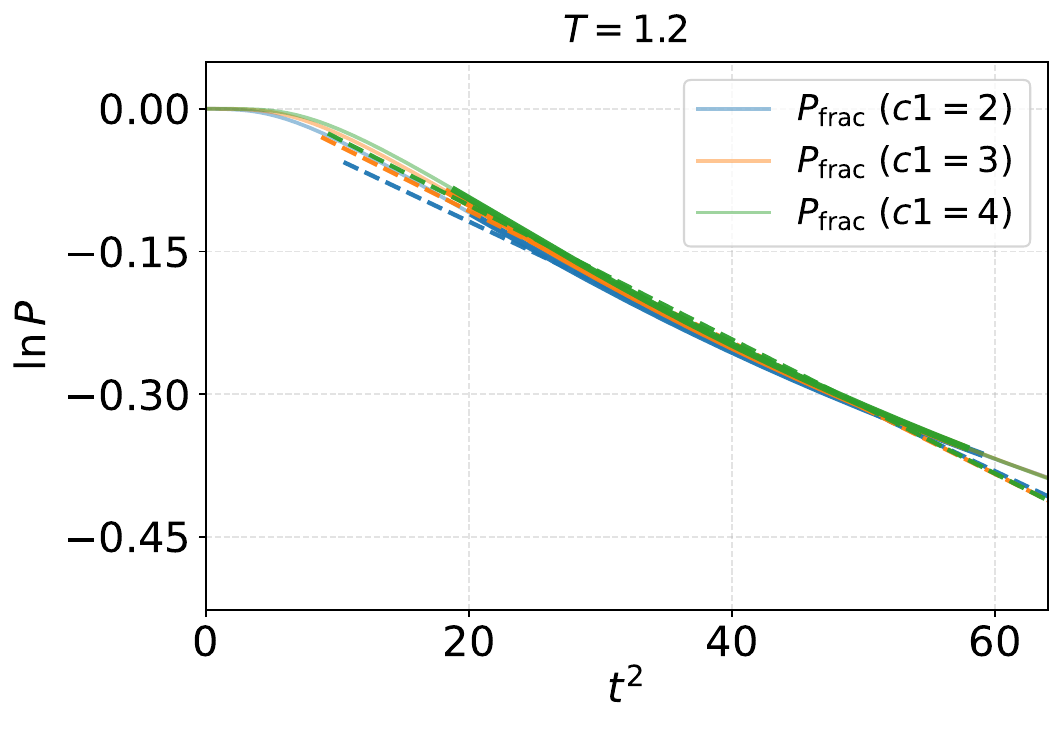}
    \includegraphics[width=0.22\linewidth]{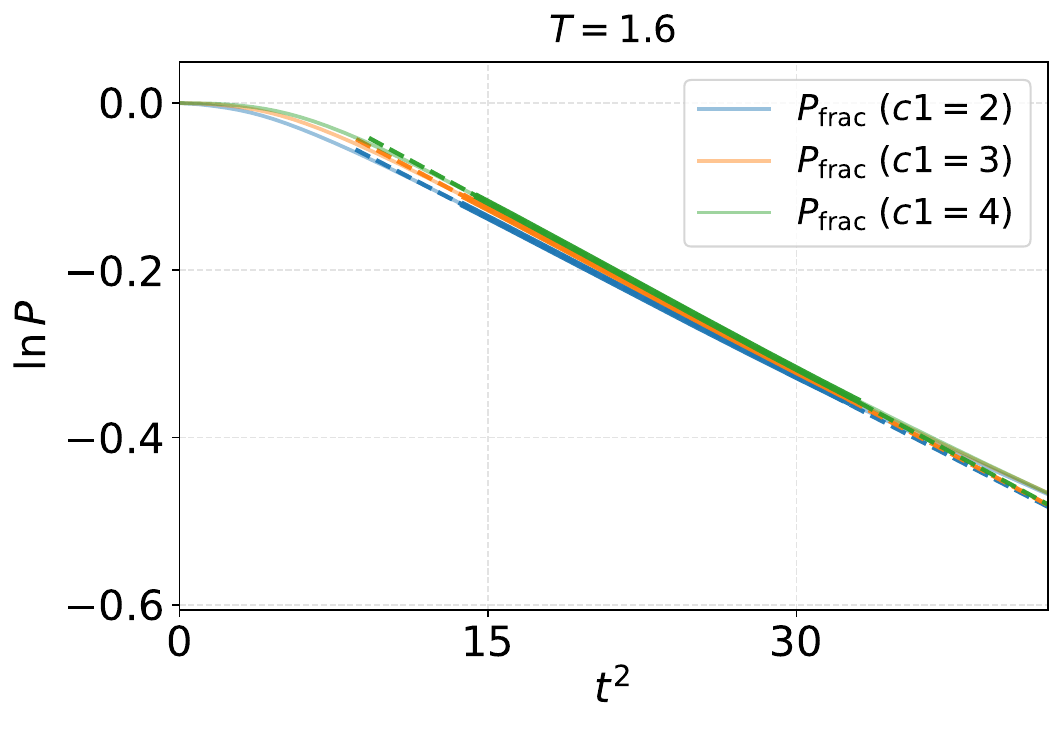}
    \caption{Time evolution of the fraction-based observable for different coarse-graining lengths. Here $c_2=0.8$, $c_3=5$, and $c_4=3$.}
    \label{fig:app_c1_pfrac}
\end{figure}

We next examine the threshold offset $c_2$. Since $c_2$ enters directly through the field-value condition $\phi_\ell>\phi_{\rm th}$, it can affect the time at which a given configuration is counted as transformed. The comparison in Figs.~\ref{fig:app_c2_psurv} and \ref{fig:app_c2_pfrac} shows that the connected-cluster criterion is relatively stable under threshold variations compared with the fraction-based and global-average observables. This behavior is expected. Once a genuinely supercritical domain has formed, most field values inside the domain lie well on the true-vacuum side, so moderate changes of the threshold have little effect on the connected-domain classification. By contrast, the fraction-based observable and the global average are more sensitive to marginal field values near the threshold, especially at higher temperature where local oscillations and multiple seeds become more frequent.

\begin{figure}[H]
    \centering
    \includegraphics[width=0.22\linewidth]{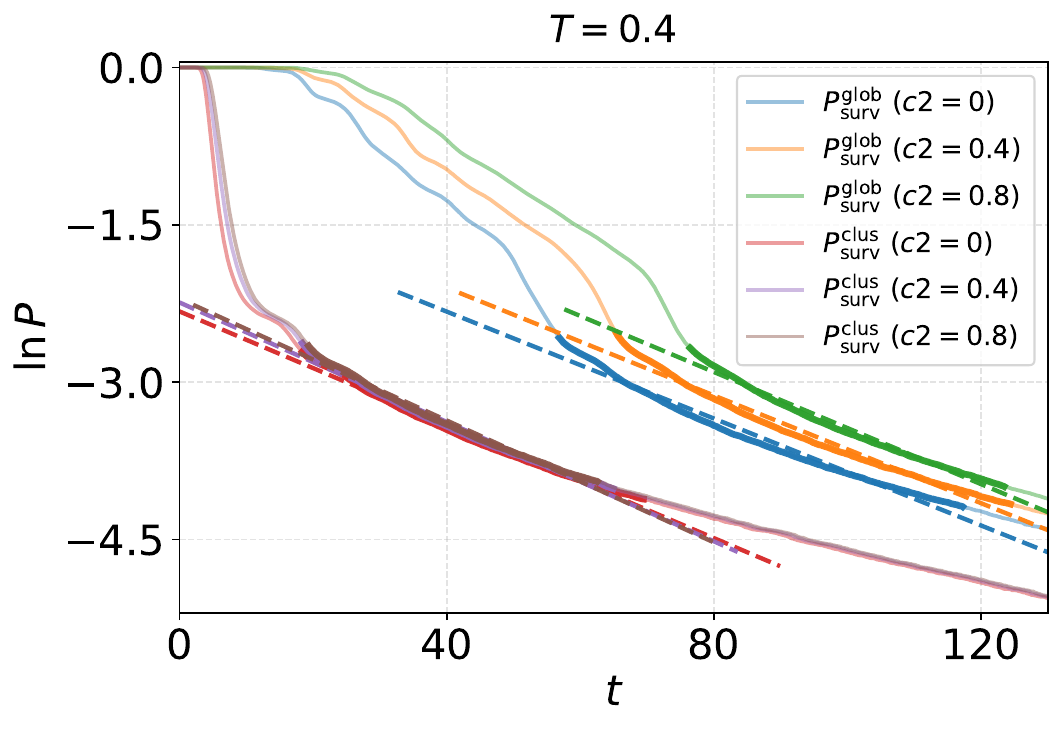}
    \includegraphics[width=0.22\linewidth]{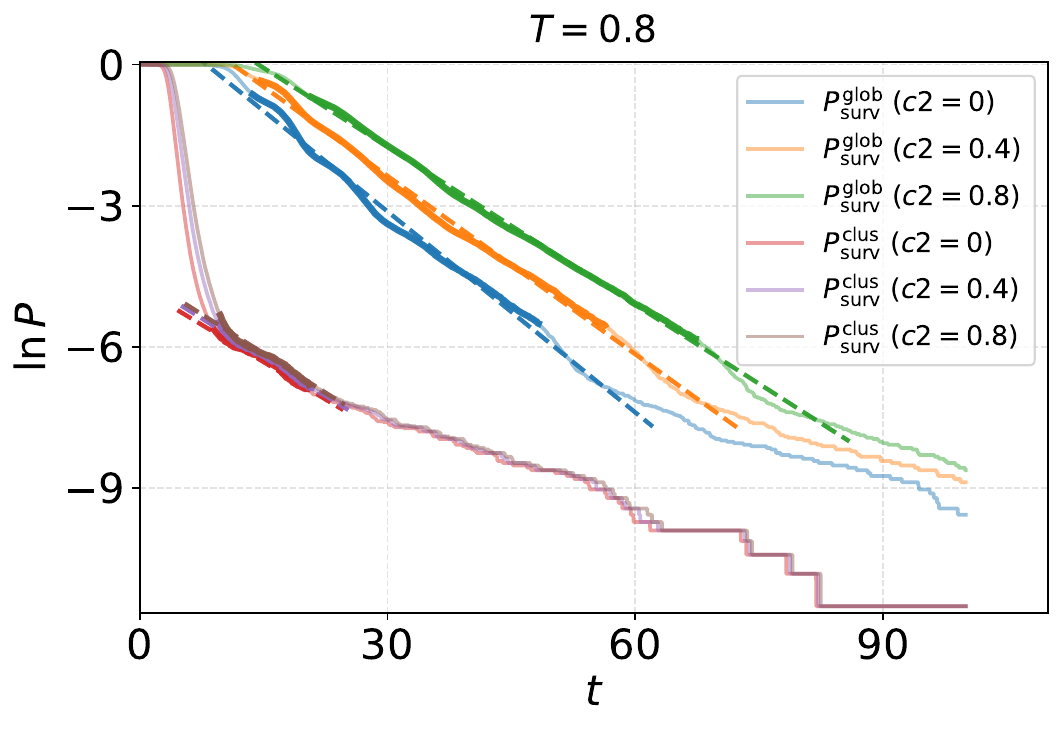}
    \includegraphics[width=0.22\linewidth]{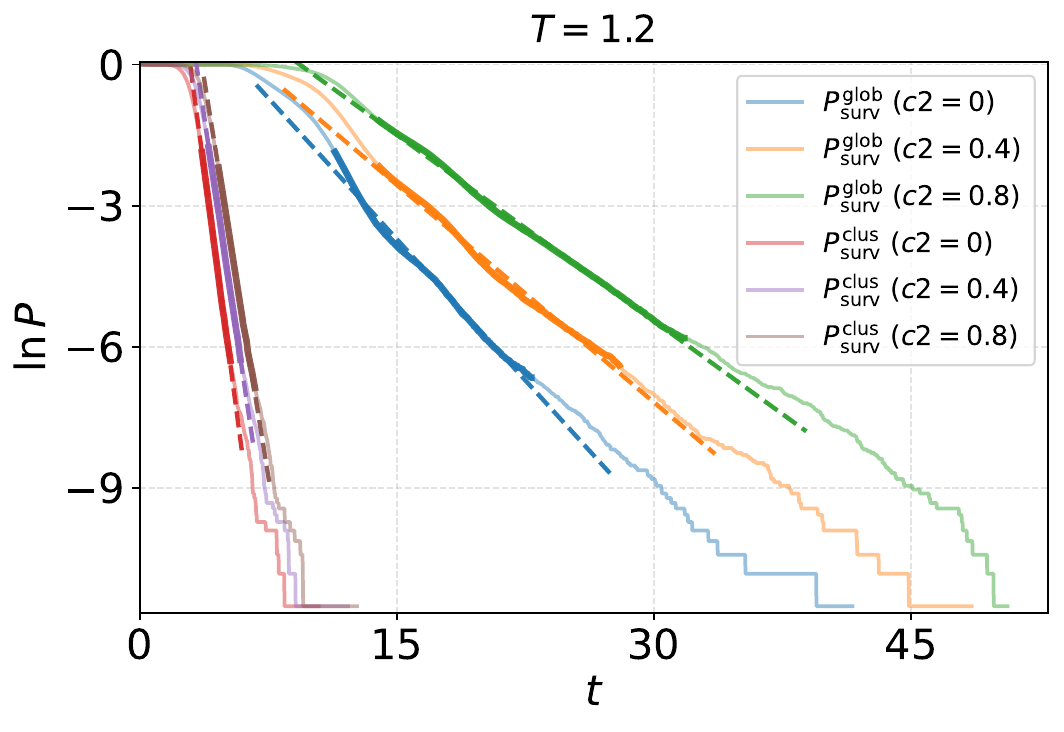}
    \includegraphics[width=0.22\linewidth]{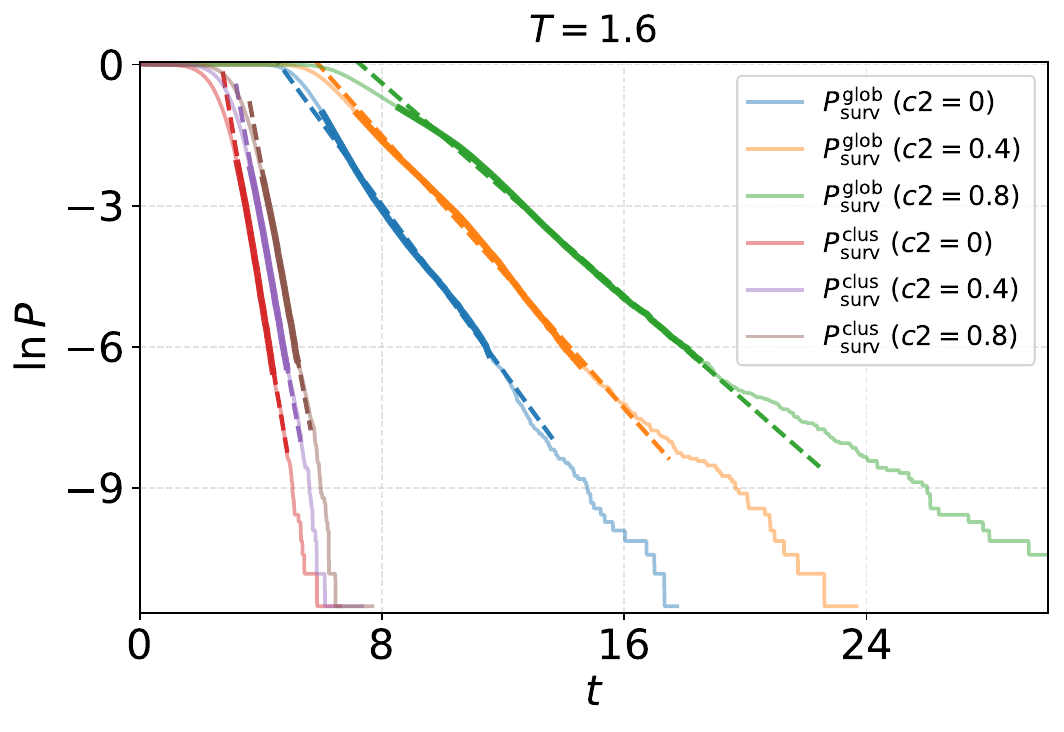}
    \caption{Time evolution of the survival probability for different threshold offsets. Here $c_1=3$, $c_3=5$, and $c_4=3$.}
    \label{fig:app_c2_psurv}
\end{figure}

\begin{figure}[H]
    \centering
    \includegraphics[width=0.22\linewidth]{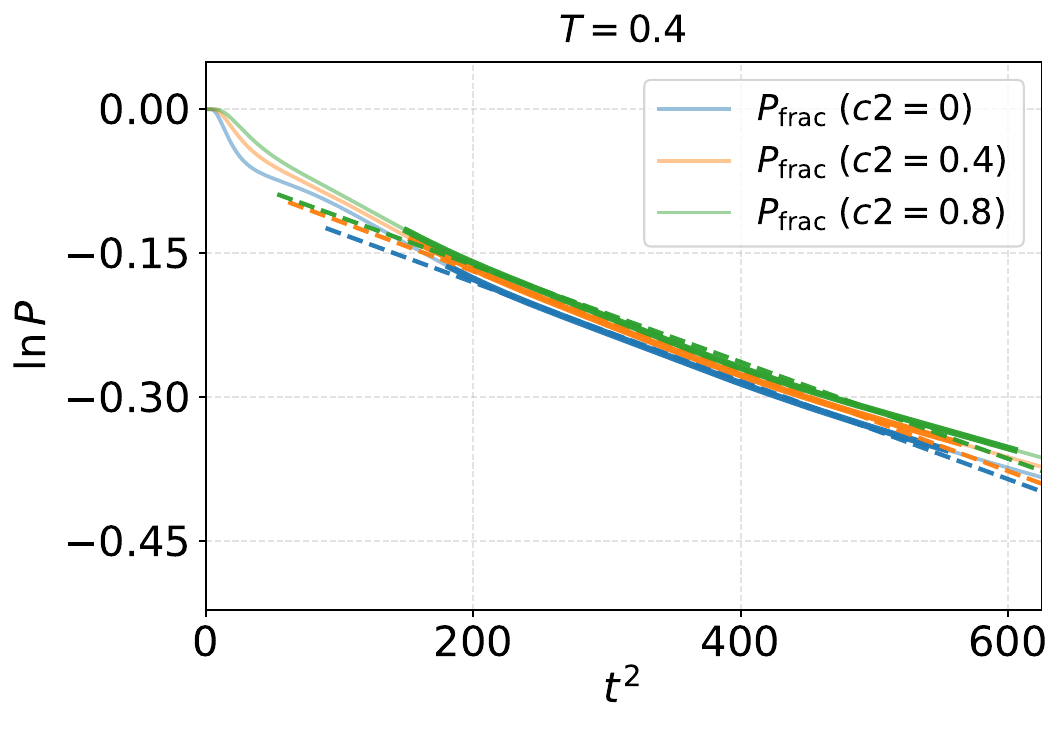}
    \includegraphics[width=0.22\linewidth]{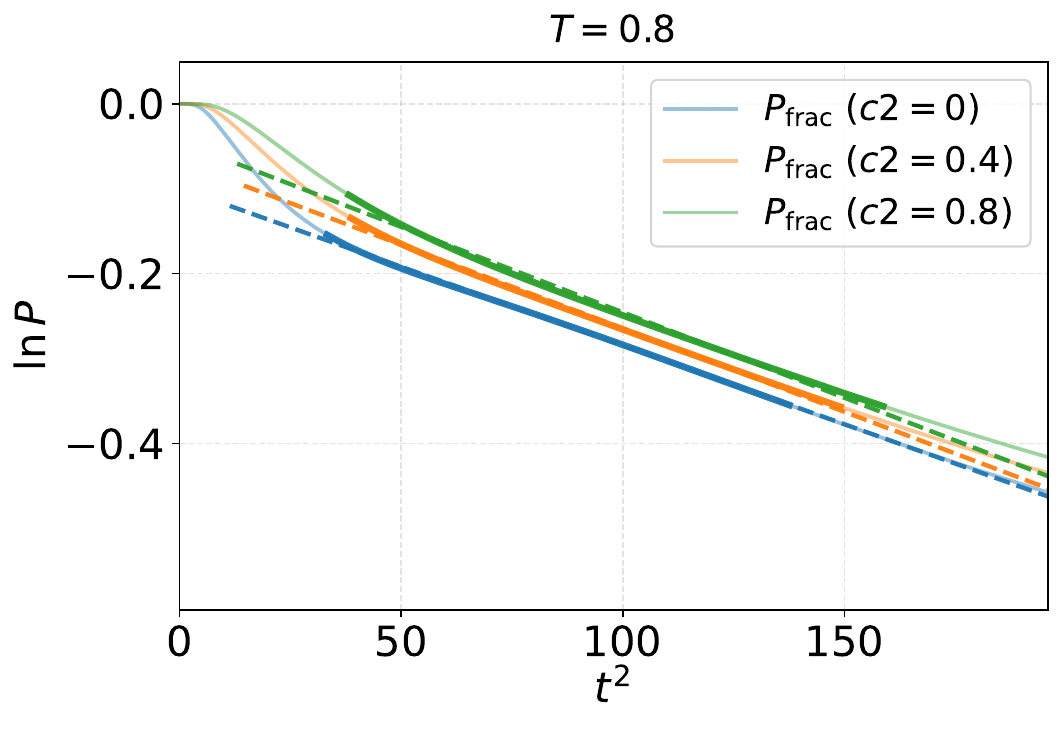}
    \includegraphics[width=0.22\linewidth]{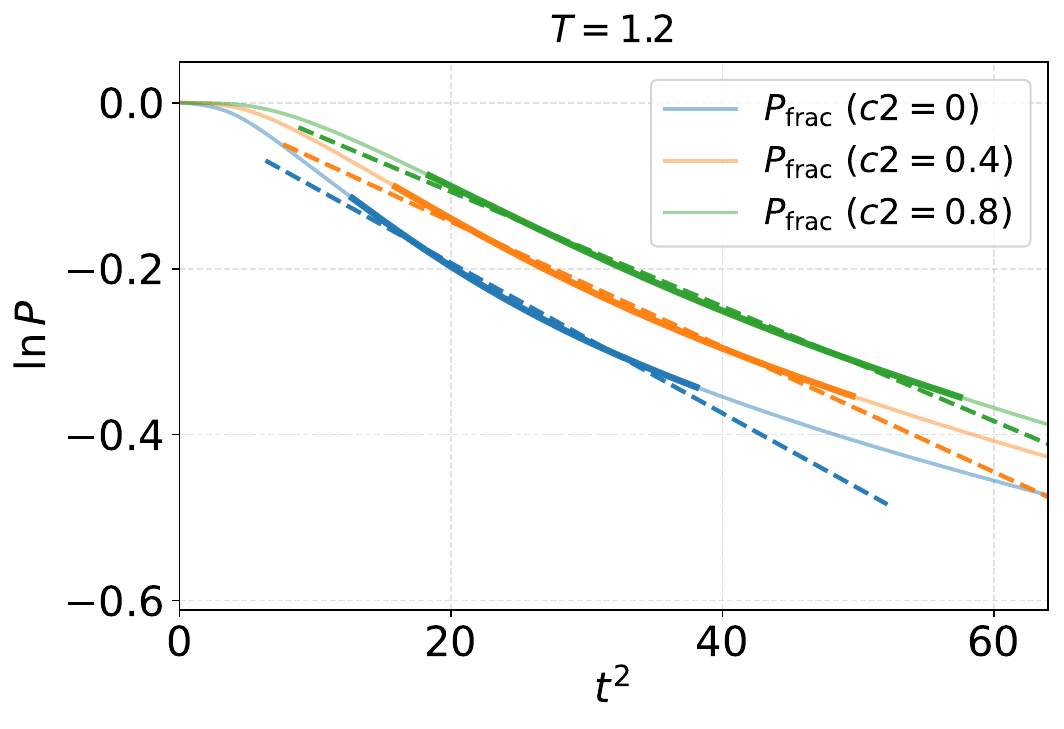}
    \includegraphics[width=0.22\linewidth]{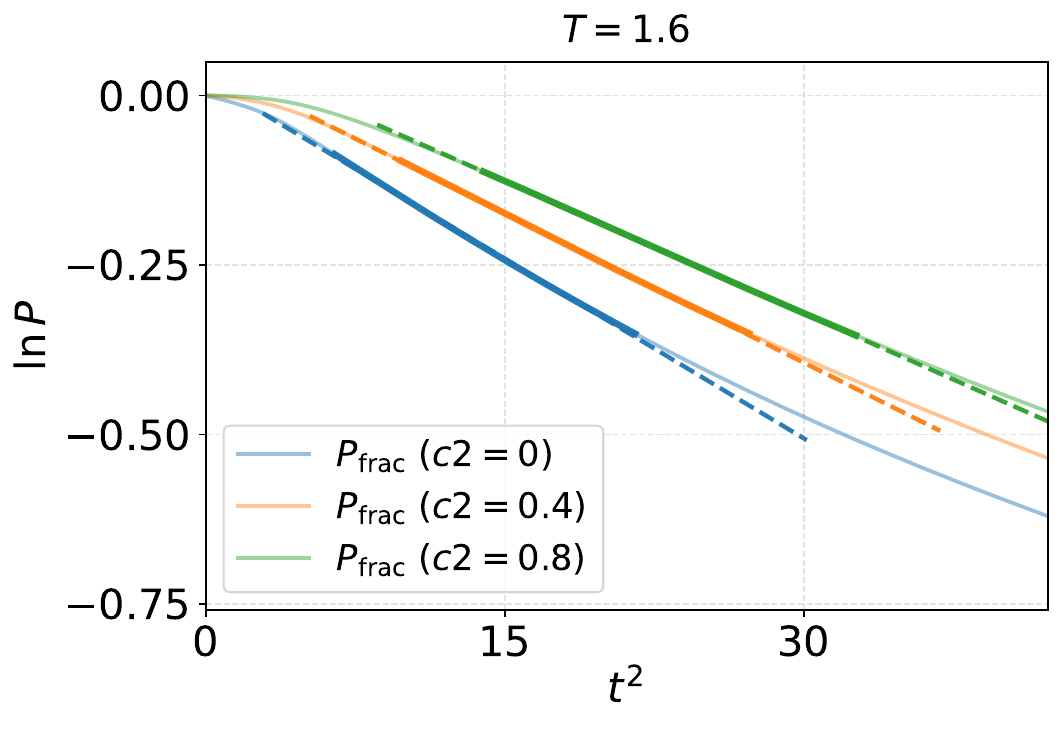}
    \caption{Time evolution of the fraction-based observable for different threshold offsets. Here $c_1=3$, $c_3=5$, and $c_4=3$.}
    \label{fig:app_c2_pfrac}
\end{figure}

The parameter $c_3$ controls the critical size of a connected supercritical interval. This parameter is especially important because it determines the minimum spatial scale required for a local transformed region to be interpreted as a genuine supercritical domain. Fig.~\ref{fig:app_c3_psurv} shows the connected-cluster survival probability for different values of $c_3$. At low temperature, changing $L_c$ has only a weak effect, because the decay is closer to a single bubble dominated process and the relevant connected domain, once formed, is typically large enough to pass a broad range of $L_c$ choices. At higher temperature, multiple seeds, local collisions, and transient connected domains become more common, so the detailed value of $L_c$ has a larger impact on the classification of decay events. Increasing $L_c$ makes the connected-cluster criterion more restrictive and can move the survival curve toward the global-average result in some temperature ranges. However, this behavior is not universal: at sufficiently high temperature, a larger $L_c$ may instead delay the cluster-based decay assignment. Thus $L_c$ should be viewed as a probe of spatial connectivity and persistence, not as a parameter that simply interpolates between local and global definitions of decay.

\begin{figure}[!htp]
    \centering
    \includegraphics[width=0.8\linewidth]{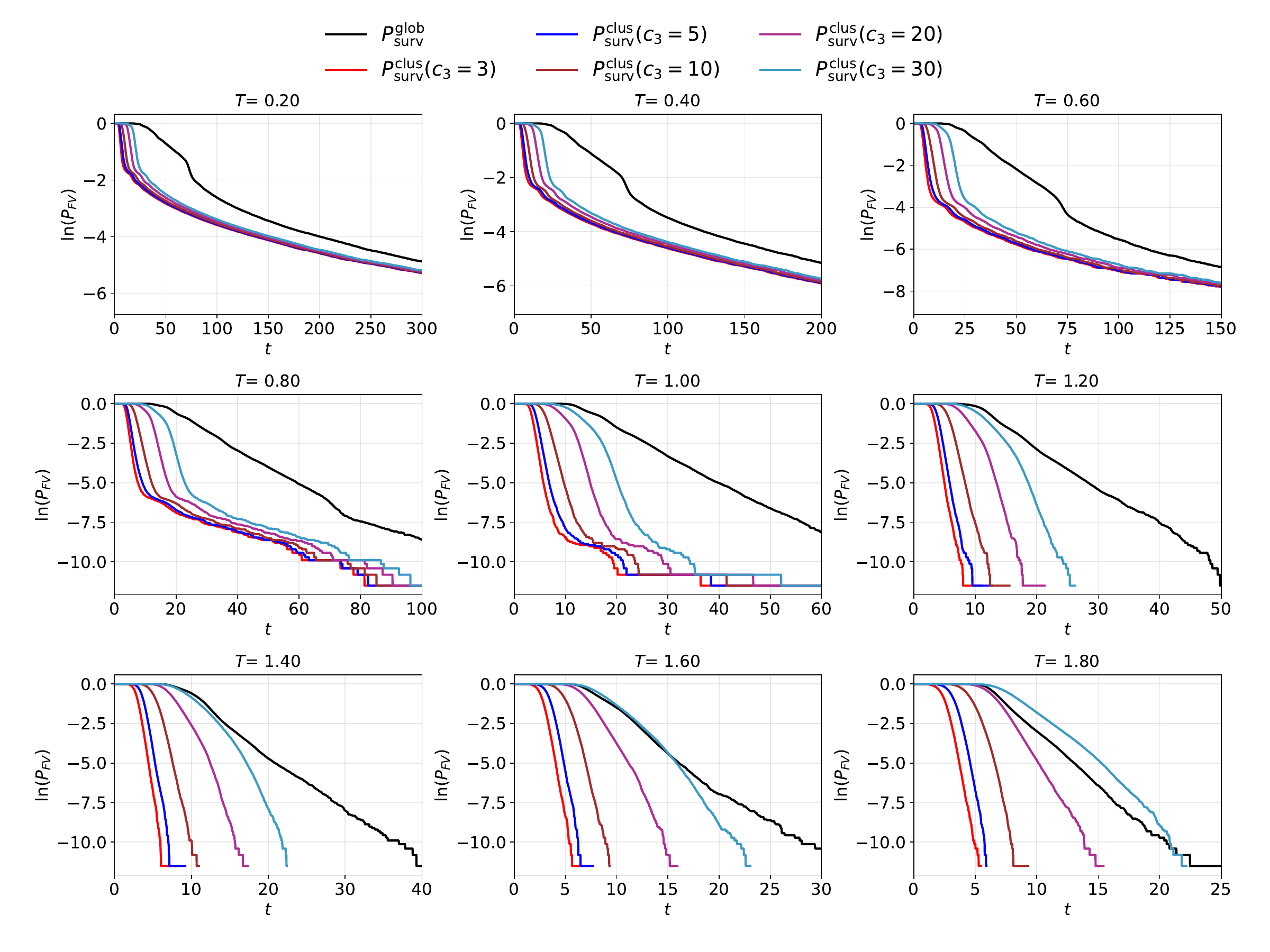}
    \caption{Time evolution of $P_{\rm surv}^{\rm clus}$ for different critical lengths. Here $c_1=3$, $c_2=0.8$, and $c_4=3$.}
    \label{fig:app_c3_psurv}
\end{figure}

Finally, we examine the holding time coefficient $c_4$. The purpose of $\tau_{\rm hold}$ is to reject transient threshold crossings and short lived connected intervals caused by local oscillations. For the values used here, $c_4=2,3,5$, Fig.~\ref{fig:app_c4_psurv} shows that the connected-cluster survival curves are almost indistinguishable. Thus, within this moderate range, the decay classification is controlled mainly by the threshold and connected length requirement, while the additional holding-time condition only confirms the persistence of domains that have already been identified as connected. The benchmark choice used in the main text is therefore stable under the tested variation of $c_4$. 
% Exploring much larger holding times would require a separate scan, preferably calibrated to the bubble size or local oscillation time scale.

\begin{figure}[!htp]
    \centering
    \includegraphics[width=0.8\linewidth]{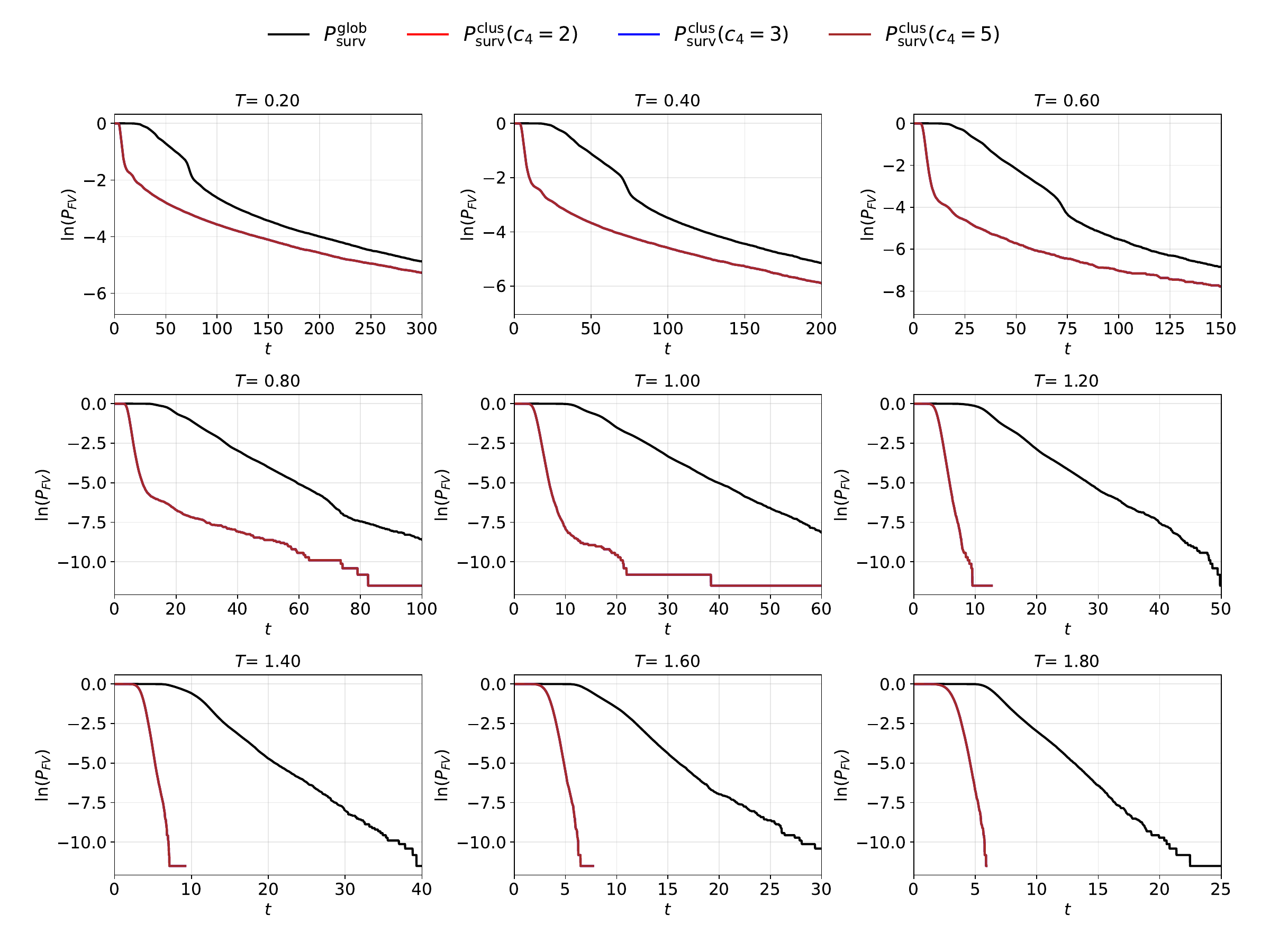}
    \caption{Time evolution of $P_{\rm surv}^{\rm clus}$ for different holding times. Here $c_1=3$, $c_2=0.8$, and $c_3=5$.}
    \label{fig:app_c4_psurv}
\end{figure}

Overall, the parameter scans show that the detailed early-time behavior of the real-time observables can depend on the operational definition of $\Theta_{FV}$, as expected. However, the extracted decay rates and their temperature dependence are stable under reasonable variations of $c_1,c_2,c_3$, and $c_4$ once the fitting-window prescription described above is applied. The connected-cluster criterion is particularly useful because it directly targets the formation of persistent supercritical domains, while the fraction-based observable provides a complementary Avrami-type diagnostic of spatial conversion and domain growth.

\section{Effective temperature from low-momentum spectra}
\label{app:Teff}

The input temperature $T_{\rm ini}$ specifies the thermal part of the initial Hartree-Gaussian Wigner distribution. However, the sampled ensemble also contains zero-point fluctuations, which dominate the high momentum part of the spectrum. Since false vacuum decay is controlled primarily by long wavelength field configurations, it is useful to introduce an effective temperature $T_{\rm eff}$ that characterizes the low momentum thermal sector after the initial short time adjustment.

We extract $T_{\rm eff}$ from the ensemble-averaged momentum spectrum. In the Rayleigh-Jeans regime, the momentum spectrum satisfies
\begin{equation}
{\cal P}_{\Pi}(k,t)\simeq T_{\rm eff}(t)
\end{equation}
for sufficiently low momentum modes. We therefore average ${\cal P}_{\Pi}(k,t)$ over a low-$k$ window,
\begin{equation}
T_{\rm eff}(t)=
\frac{1}{N_k}
\sum_{0<k<k_{\rm cut}}{\cal P}_{\Pi}(k,t),
\label{eq:Teff_app}
\end{equation}
where $N_k$ is the number of modes included in the average. The cutoff $k_{\rm cut}$ is chosen to be of order the inverse critical-bubble scale. In the present potential, the static critical bubble radius is approximately $R_c\simeq 2.44$, and the cutoff is chosen accordingly. We then identify an early time plateau of $T_{\rm eff}(t)$ before substantial decay occurs and use the plateau value as the effective temperature associated with the run.

In the main analysis we use the momentum spectrum rather than the field spectrum, because ${\cal P}_{\Pi}(k,t)$ is less sensitive to the slow drift of the mean field and to large scale inhomogeneous domain formation. We also use the ensemble-averaged spectrum over all samples in the early pre-decay stage, rather than averaging only over undecayed samples within the decay fitting window as in Ref.~\cite{Pirvu:2023plk}, because the latter choice would introduce an additional dependence on the decay criterion itself. This prescription corresponds to the scheme used in the main text.

Figure~\ref{fig:Teff_methods_app} compares several prescriptions for extracting $T_{\rm eff}$ from the low-momentum $\Pi$ spectrum. The ``All Avg'' result, used in the main text, is obtained by averaging the low-$k$ spectrum over all samples. Time window 1 uses a self-referenced selection based only on $T_{\rm eff}(t)$: an early-time baseline is first estimated before the noticeable rise of $T_{\rm eff}(t)$, and the final value is obtained from the stable interval just before this rise. If no clear onset is found, the earliest stable plateau is used instead. Time window 2 uses the decay-rate fitting interval and averages only over undecayed samples, providing a survival-conditioned comparison. We also compare two infrared cutoffs, $k_{\rm cut}=\pi/R_c$ and $k_{\rm cut}=\pi/L_c$, to check that the extracted temperature is not overly sensitive to whether the low-momentum window is set by the static critical-bubble scale or by the connected-cluster length scale. The zero-mode matching curve is included only as a reference, since a single mode is more sensitive to finite-volume fluctuations. The different prescriptions give similar trends, while the all-sample low-$k$ average with $k_{\rm cut}=\pi/R_c$ is used as the practical temperature diagnostic in the main analysis.

\begin{figure}[!htp]
    \centering
    \includegraphics[width=0.8\linewidth]{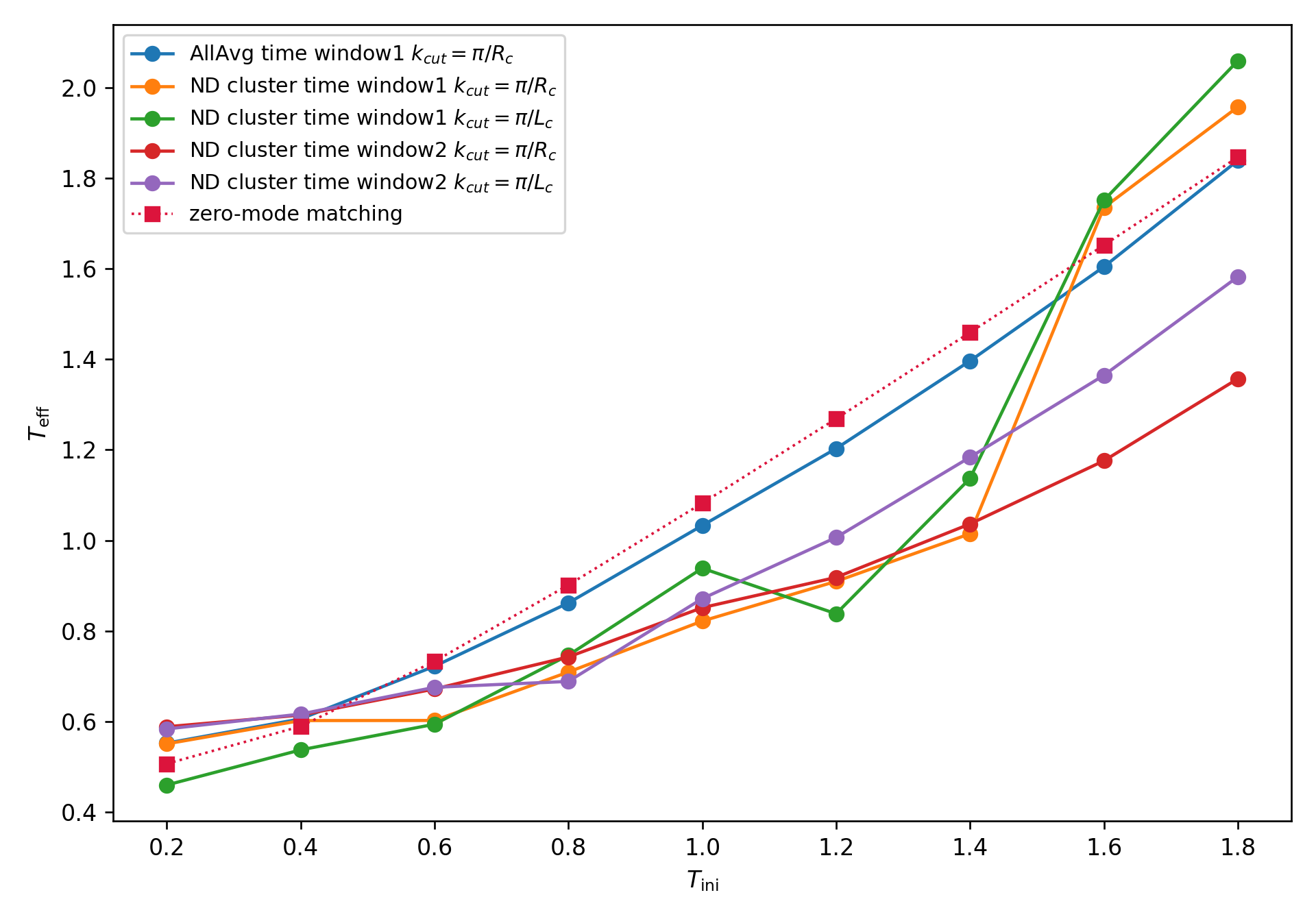}
    \caption{Effective temperature extracted from the low-momentum part of the $\Pi$ spectrum using different prescriptions. The red dotted curve denotes the zero-mode matching reference.}
    \label{fig:Teff_methods_app}
\end{figure}

Representative examples of the early-time evolution of $T_{\rm eff}(t)$ are shown in Fig.~\ref{fig:Teff_vs_time}. The shaded regions denote the time windows used to define the plateau average. At very early times, the spectrum still contains transient information from the sampled initial ensemble. At later times, the growth of inhomogeneous domains and the onset of decay can distort the low-momentum spectrum. We therefore use an early time window, before substantial decay occurs, to characterize the effective thermal background of each run.

As shown in Fig.~\ref{fig:Teff_vs_time}, $T_{\rm eff}(t)$ exhibits a mild oscillatory relaxation during the early stage. The window-averaged value remains close to the temperature scale inferred from the low-momentum spectrum and provides a stable mapping from $T_{\rm ini}$ to $T_{\rm eff}$. The horizontal zero-mode matching value is included as a reference, but our main estimate is obtained by averaging over a low-momentum band and over an early-time window. This makes the extracted $T_{\rm eff}$ less sensitive to finite-volume fluctuations and to the behavior of any single momentum mode.

\begin{figure}[!htp]
    \centering
    \includegraphics[width=0.4\linewidth]{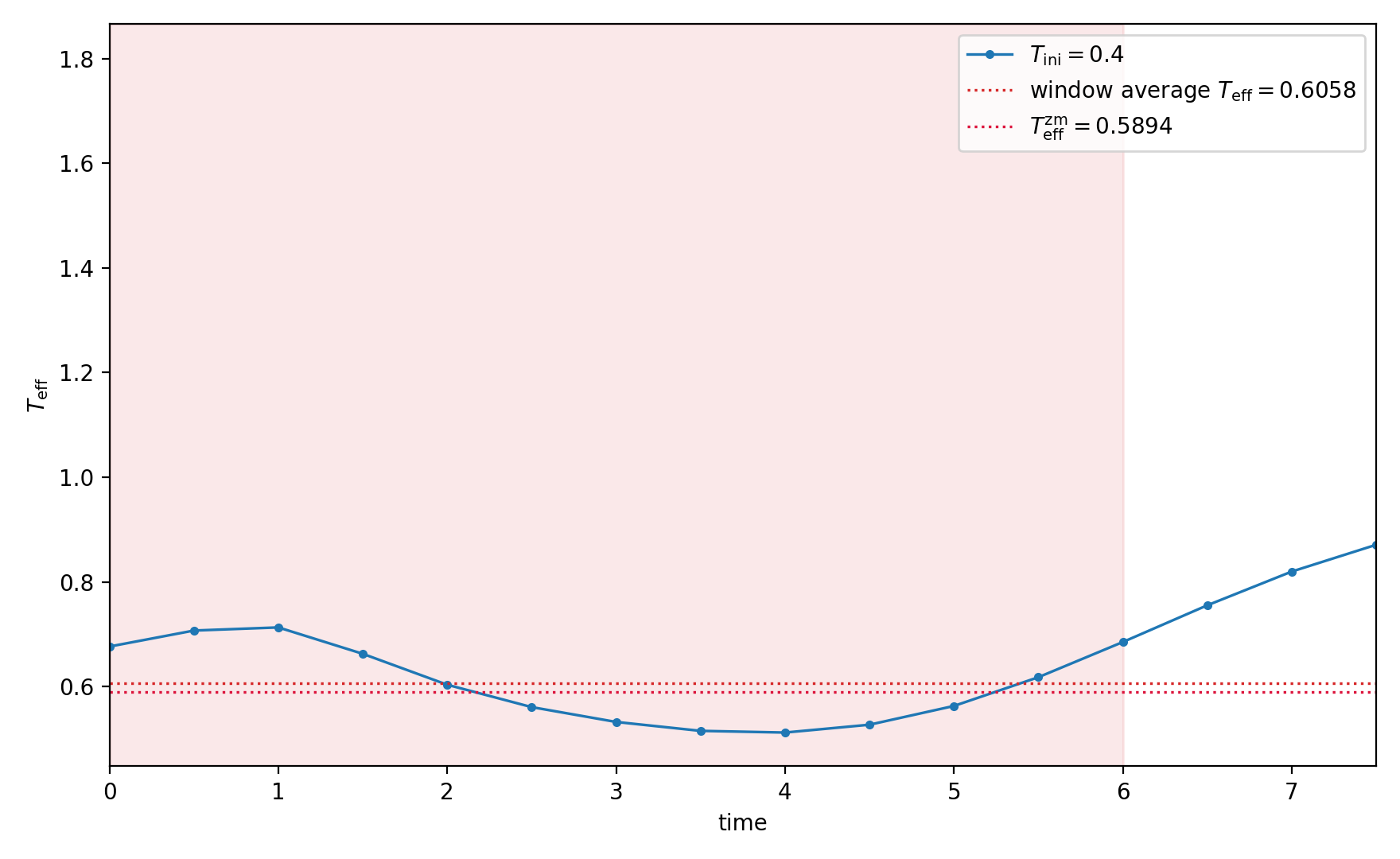}
    \includegraphics[width=0.4\linewidth]{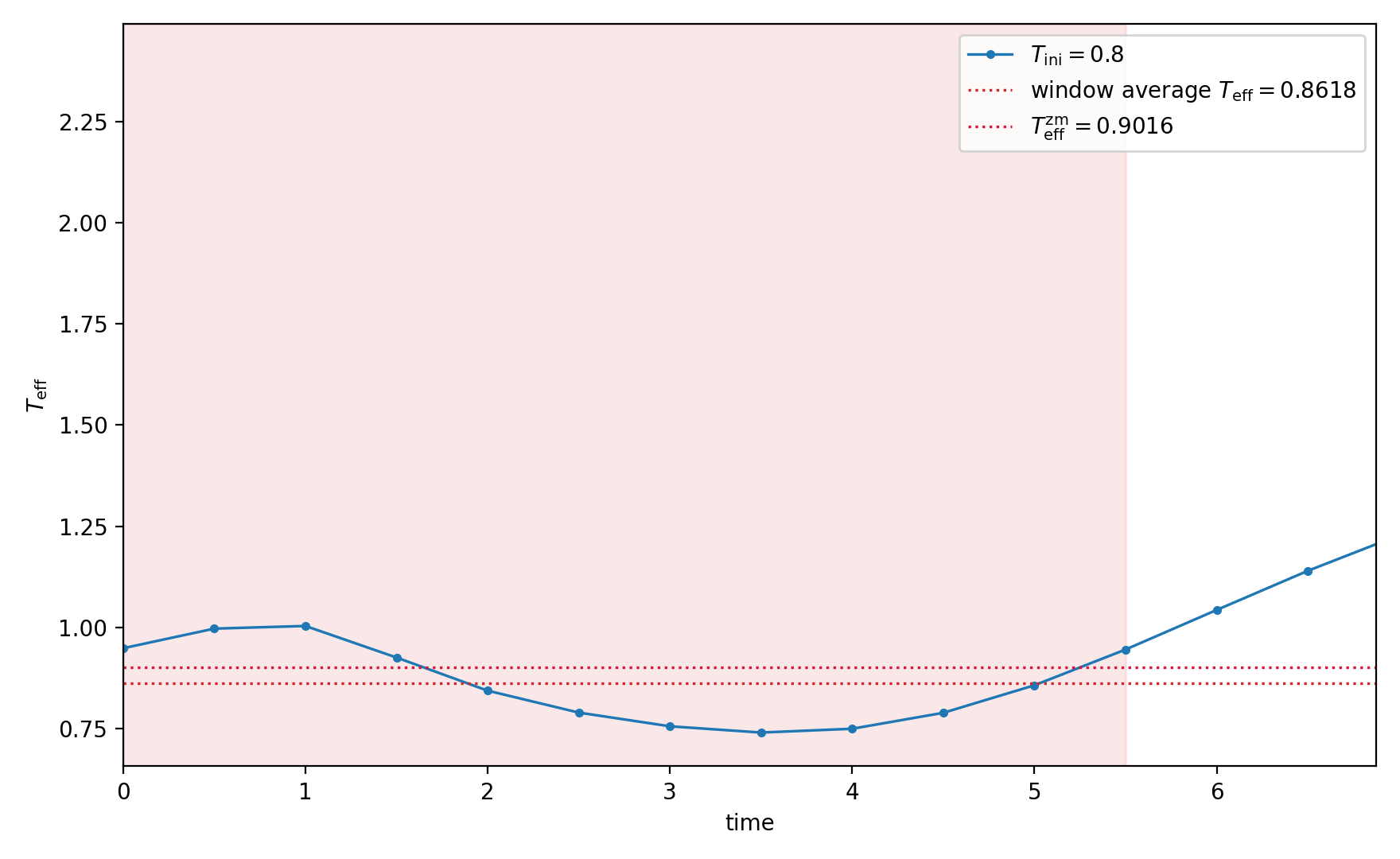}
    \includegraphics[width=0.4\linewidth]{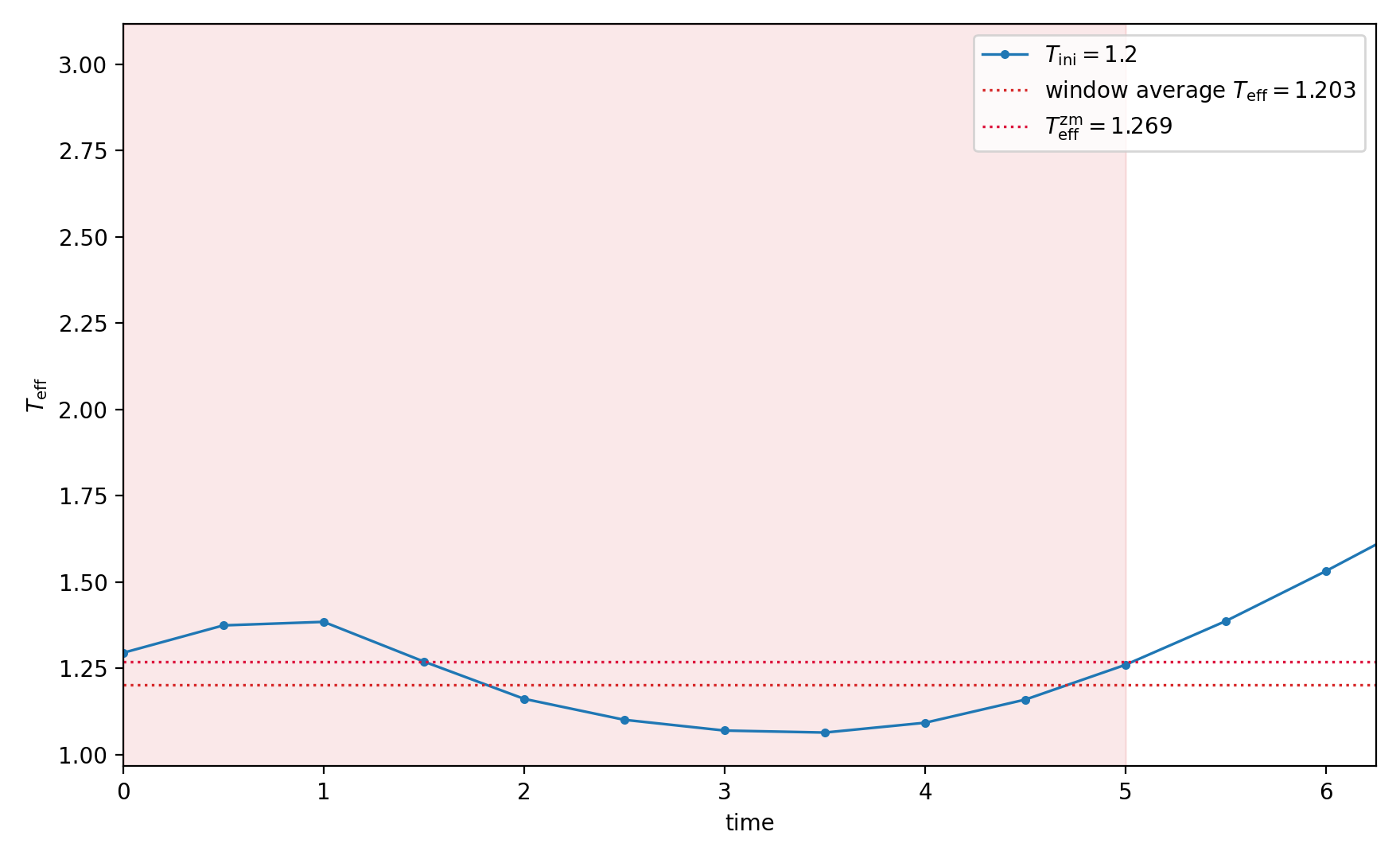}
    \includegraphics[width=0.4\linewidth]{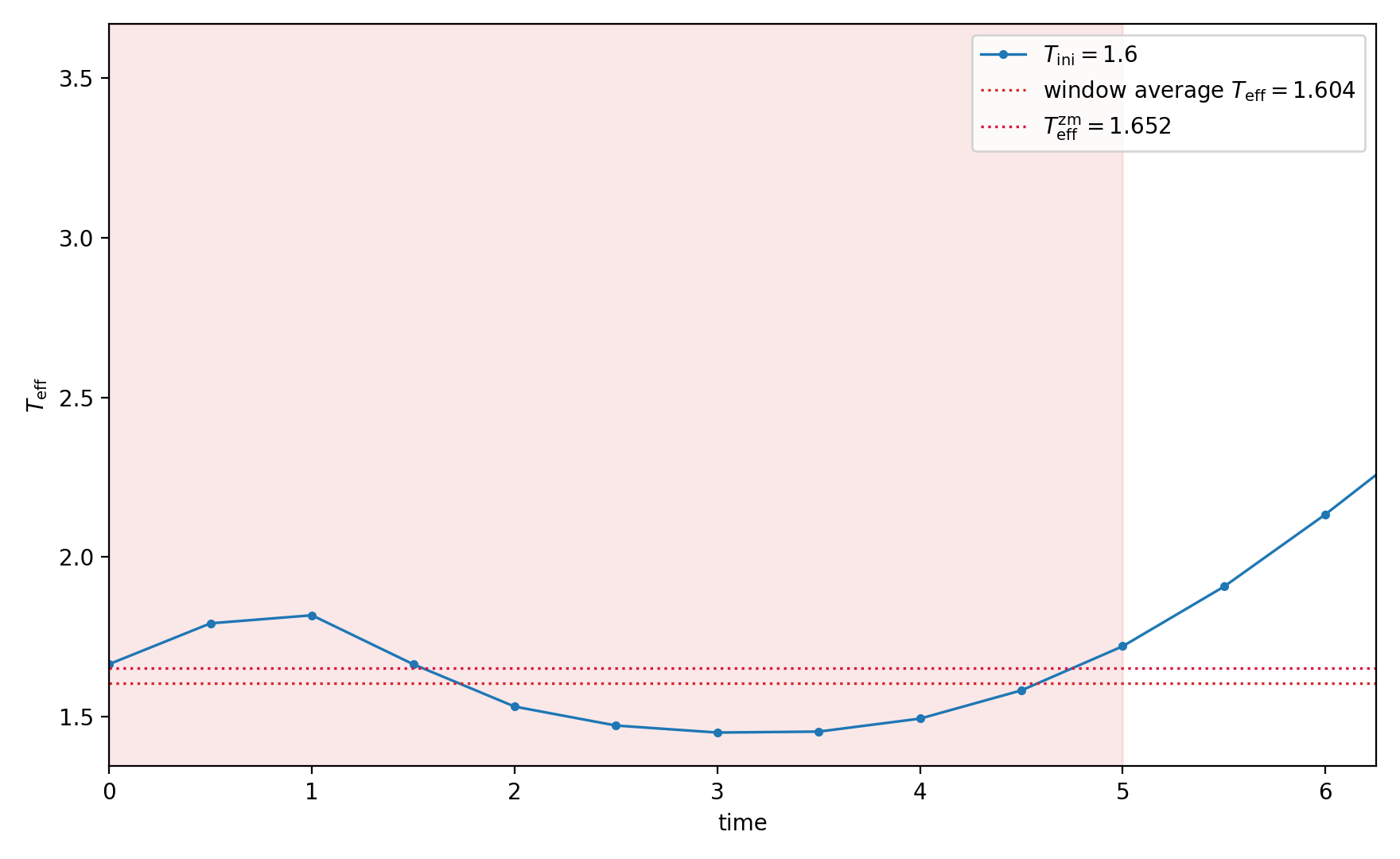}
    \caption{Early-time evolution of the effective temperature extracted from the low-momentum part of the $\Pi$ spectrum for several input temperatures. The shaded regions denote the time windows used for the window-averaged estimate of $T_{\rm eff}$. The horizontal dotted lines show the corresponding zero-mode matching reference values.}
    \label{fig:Teff_vs_time}
\end{figure}

\bibliographystyle{apsrev4-1}

\end{onecolumngrid}

\clearpage

\end{document}